\documentclass[twocolumn,showpacs,amsmath,amssymb,showkeys,
floatfix,prl]{revtex4}
\usepackage{amsfonts,amsmath}
\usepackage{graphicx}
\usepackage{dcolumn}
\usepackage{bm}

\begin{document}

\title{Implications of the Dirac CP phase upon parametric resonance for sub-GeV neutrinos}

\author{Edouard A.~Hay}
\author{David C.~Latimer}
\affiliation{Department of Physics, Reed College, Portland, Oregon 97202, USA
}
\date{\today}

\begin{abstract}
We perform an analytic and numerical  study of parametric resonance in a three-neutrino framework for sub-GeV neutrinos which travel through a periodic density profile.  Commensurate with the initial level of approximation, we develop a parametric resonance condition similar to the exact condition for two-neutrino systems.  For a castle wall density profile, the $\nu_e \to \nu_\mu$ oscillation probability is enhanced significantly and bounded by $\cos^2 \theta_{23}$.  The CP phase $\delta$ enters into the oscillation probability as a phase shift.  For several cases, we examine the interplay between the characteristics of the castle wall profile and the CP phase and determine which profiles maximize the separation between oscillations with $\delta = 0,\pm \frac{\pi}{2},\pi$.  We also  consider neutrinos which travel along a chord through the earth, passing from the mantle to core and back to mantle again.  Significant enhancement of the oscillation probability is seen even in the case in which the neutrino energy is far from the MSW resonant energies.  At 500 GeV, the difference between oscillation probabilities with $\delta=0$ and $\delta=\frac{\pi}{2}$ is maximized. 

\end{abstract}

\maketitle
\section{Introduction}

The phenomenon of neutrino oscillations is a consequence of the fact neutrino weak interaction states are superpositions of the mass eigenstates.  Relevant to  oscillation phenomenology are the neutrino mass-squared differences $\Delta_{jk} := m_j^2 - m_k^2$ and the PMNS mixing matrix which we denote as $U$ \cite{p,mns}.  The mixing matrix can be parametrized in terms of three real mixing angles $\theta_{jk}$, with $j,k =1,2,3$ and $j < k$, and the Dirac CP phase $\delta$.  The overwhelming majority of neutrino oscillation data fits quite well within this standard three-neutrino framework; however, when considering a subclass of the experiments, one can often accurately understand  this restricted data in a two-neutrino framework, requiring only a single mass-squared difference $\Delta m^2$ and mixing angle $\theta$.    Solar neutrino experiments and long baseline (LBL) reactor experiments can be approximately parameterized by $\Delta_{21}$ and $\theta_{12}$, and atmospheric and some accelerator neutrino experiments likewise can be described with $\Delta_{32}$ and $\theta_{23}$.  The ability to separate the data as such speaks to the smallness of the mixing angle $\theta_{13}$ and the ratio of mass-squared differences $|\Delta_{21}/\Delta_{32} |$.

Analogies exist between neutrino oscillations and mechanical oscillations, and in particular,
mechanical oscillators can exhibit large amplitude oscillations when some of the oscillation parameters change periodically.  This parametric resonance is particularly prominent when the parameters change at twice the natural frequency of oscillation.  Examples of parametric resonance are pendula with vertically oscillating supports \cite{ll_mechanics} and Faraday waves, surface instabilities created in a vertically oscillating container of fluid \cite{faraday}.  The possibility of parametrically enhanced neutrino oscillations was first noted in Refs.~\cite{ermilova,akhmedov_1}; it was shown that if neutrinos travel through matter with a particular periodic density profile the oscillation probability can be considerably enhanced.  

As neutrinos travel through matter, the mixing angles and mass-squared differences are effectively modified as described by the MSW effect \cite{ms,w}.  Neutrinos which propagate long distances through matter of sufficient densities can incur significant interactions which are diagonal in
flavor, as the interactions are mediated by the charged and neutral currents of the weak interaction.    Neutral current interactions are democratic amongst the flavors, leaving the oscillation probabilities unchanged; however, since ordinary matter consists of electrons, protons and neutrons, charged current interactions affect only the electron (anti-)neutrinos, modifying the oscillation probability.  The upshot is that  the neutrino oscillation parameters effectively change in a periodic manner if the neutrinos travel through matter with a periodic density profile, leading to possibility of parametric enhancement.

Parametric resonance in neutrino oscillations has been studied extensively through both analytical and numerical means \cite{ermilova,akhmedov_1,krastev,liu_smirn, liu_mikh_smirn,petcov_param,akh_long,akh_atmos_SK,floquet,chizhov,kimura,param_highE, akh_13, akh_12, akh_nu_osc_cpv,param_fourier}.    
Given the small mass-squared differences of the neutrinos and the available energies from high flux sources, it is not possible to set up a tabletop experiment with the appropriate density profile so as to demonstrate parametric resonant oscillations; however, in Refs.~\cite{liu_smirn, liu_mikh_smirn}, it was realized that the density profile of earth's interior might provide a suitable laboratory.  Indeed, the earth's density may be approximated as piecewise constant with two main regions--a mantle surrounding a denser core \cite{prem}.  A periodic potential consisting of two piecewise constant regions of differing densities is often referred to as a ``castle wall" potential.  Exact analytic solutions for two-neutrino oscillations through such castle wall profiles exist and serve as a fundamental tool for understanding parametric enhancement for core-crossing trajectories
\cite{akhmedov_1,akh_long,floquet}.   Specific applications consider atmospheric neutrinos which travel through the earth \cite{akh_atmos_SK,param_highE,akh_13, akh_12,akh_nu_osc_cpv,param_fourier}.   Relatively exhaustive semi-analytic and numerical studies for neutrino oscillations in the earth were done in Refs.~\cite{akh_13, akh_12} where resonance regions are shown to follow from generalized amplitude and phase conditions.  
For semi-analytic treatments of three-neutrino oscillations, one typically reduces the problem to an effective two-neutrino system at varying levels of approximation.  In this manner, one may incorporate the Dirac CP phase into the analysis, something not possible in a pure two-neutrino theory.  This is what is done in Refs.~\cite{akh_12,akh_nu_osc_cpv} where the consequences of CP violation to the oscillation probability is considered for neutrinos with energies in excess of 1 GeV traveling through the earth; particular attention is paid to the interference between oscillations due to the $\Delta_{21}$ and $\Delta_{31}$ mass-squared differences. 

We will examine, herein, the interplay between CP violating effects and parametric resonance for sub-GeV neutrinos.  Using approximations relevant for sub-GeV neutrinos and mantle/core densities, the three neutrino system can be cast into an effective two-neutrino system via the so-called propagation basis \cite{peres_2004}.  We apply existing work on two-neutrino parametric resonance  to a novel semi-analytic study of sub-GeV neutrinos, including CP violation.  At a level of approximation commensurate with that used to rotate to the propagation basis, we find a condition for parametric resonance similar to the two neutrino case, and upon implementing this condition, we are able to assess the value of the $\nu_e \to \nu_\mu$ oscillation probability at the end of an integer number of periods of the matter potential.   We show that $\mathcal{P}_{e\mu}$ is enhanced and bounded by $\cos^2 \theta_{23}$ here.   Also, we are able to determine the characteristics of a castle wall profile that will lead to maximum separation between $\mathcal{P}_{e\mu}$ curves employing different values of the CP phase $\delta=0,\pm\frac{\pi}{2},\pi$.   We then turn briefly to the situation in which neutrinos do not travel through an integer number of periods as this is relevant for atmospheric neutrinos passing through the earth.  Again, we focus upon parameters which implement parametric resonance and then look at oscillations with different values of the CP phase.  

Our focus upon sub-GeV neutrinos is motivated by previous work which has shown that for long baselines the $\nu_e \to \nu_\mu$ oscillations driven by the solar mass-squared difference contain relatively sizable terms proportional to $\sin \theta_{13}$ which are CP odd \cite{peres_2002,peres_2004,dcl_th13,dcl_onechannel}.
As we enter an era of precision neutrino experiments, evidence for a nonzero value of $\theta_{13}$ mounts and with it the possibility of measuring the level of CP violation, if any, in the neutrino sector.  Strict upper bounds on the magnitude of $\theta_{13}$ were initially established by the CHOOZ reactor experiment \cite{chooz}; however, a recent reevaluation of the reactor neutrino flux \cite{reactor_flux} has somewhat relaxed this upper bound.  Furthermore, hints of nonzero $\theta_{13}$ come from  joint analyses of solar neutrino and KamLAND data \cite{balantekin,fogli_prl, maltoni_hints}.  Though statistically less significant, analyses of atmospheric neutrino experiments also favor a nonzero reactor mixing angle \cite{fogli_prog,jesus_prl,jesus_prc_th13}.    Accelerator $\nu_\mu \to \nu_e$ appearance experiments MINOS \cite{minos_prl_th13} and T2K \cite{t2k_th13} have both detected electron neutrinos above the expected background, further evidence for non-zero $\theta_{13}$.   A global analysis of this neutrino data, excluding recent reactor experiments, indicates a value of $\theta_{13}$ differing from zero by more than $3 \sigma$ \cite{fogli_2011}.  Perhaps most significant are the data from  two reactor $\bar{\nu}_e$ disappearance  experiments; both Daya Bay \cite{dayabay} and RENO \cite{reno} report nonzero values of $\sin^2 2\theta_{13}$ at the 5-$\sigma$ level.

\section{Oscillation in matter}
The ultrarelativistic limit of the evolution equation for a neutrino of energy $E$ is
\begin{equation}
i \partial_t \nu = \frac{1}{2E} U \mathcal{M} U^\dagger \nu \, ,
\end{equation}
where we define the matrix $\mathcal{M} = 
\mathrm {diag} (0, \Delta_{21}, \Delta_{31})$.    We employ the parametrization used in Ref.~\cite{peres_2004}
\begin{equation}
U = U_1(\theta_{23}) D_\delta U_2(\theta_{13}) U_3(\theta_{12}) \, ,
\end{equation}
where $U_j(\theta)$ is a proper rotation by angle $\theta$ about the $j$-th axis and $D_\delta = \mathrm{diag}(1,1,e^{i\delta})$; this is different from, but equivalent to, the standard parametrization found in Ref.~\cite{pdg}. 

When neutrinos travel through matter, the Hamiltonian accrues an effective potential due to the coherent forward scattering of the neutrinos upon electrons, protons, and neutrons which comprise the matter \cite{ms,w}.  
We include this effective potential in the evolution equation 
\begin{equation}
i \partial_t \nu = \left[ \frac{1}{2E} U \mathcal{M} U^\dagger + \mathcal{V}(x) \right]\nu  . \label{mswev}
\end{equation}
Neglecting the (irrelevant for oscillations) neutral current interaction, the operator $\mathcal{V}(x)$ exclusively acts on the electron flavor  with a magnitude $V= \sqrt{2} G_F N_e(x)$, where $G_F$ is the Fermi coupling constant  and $N_e$ is the electron number density.   We note that for anti-neutrinos, we need to change the algebraic sign of this 
potential and the CP phase $\delta$.  We shall consider only neutrinos below.

For sub-GeV neutrinos traversing the earth, 
matter effects are most easily addressed
in the propagation basis developed in Ref.~\cite{peres_2004}.  
We will briefly review this derivation.  
As the  $U_1(\theta_{23})$ portion of the mixing matrix commutes with $\mathcal{V}$, we
may rewrite the evolution equation
\begin{equation}
i \partial_t \nu' =
\left[ \frac{1}{2E} U_3(\theta_{12}) \mathcal{M}
U_3(\theta_{12})^\dagger + U_2(\theta_{13})^\dagger \mathcal{V} U_2 
(\theta_{13}) \right] \nu'
\end{equation}
with $\nu' = U_2(\theta_{13})^\dagger D_\delta^\dagger U_1(\theta_{23})^\dagger \nu$.

By conjugating the Hamiltonian in this basis via a locally defined 
$U_2(\theta)$, this
new propagation basis can be approximately described
by a Hamiltonian $\tilde{H}$ which is block diagonal.  This correction to
$\theta_{13}$ is given by
\begin{equation}
\tan {2 \theta} = \frac{2 \sin{2\theta_{13}}E V}{\Delta_{31} -s_{12}^2
\Delta_{21}
- 2\cos{2\theta_{13}}EV} \label{theta}
\end{equation}
where we use the shorthand $s_{12} := \sin \theta_{12}$.
The density of the earth's interior has an upper bound around 13 g/cm$^3$ \cite{prem}; this results in a maximum effective potential $V \sim5 \times 10^{-13}$ eV.  
As $\Delta_{31}\sim 2.4 \times 10^{-3}$ eV$^2$,  the mass-squared difference is the dominant term in the denominator of Eq.~(\ref{theta}); for $E\sim 1$ GeV, one has $\epsilon:=2EV/\Delta_{31} < 0.4$.  For sub-GeV energies,  one may approximate Eq.~(\ref{theta}) as
\begin{equation}
\theta \simeq \frac{ \sin{2\theta_{13}}E V}{\Delta_{31}}.
\end{equation}
This correction results in a modified mixing angle
\begin{equation}
\tilde \theta_{13} = \theta_{13} + \theta.
\end{equation}
With this additional rotation, we define locally the propagation basis with
$\tilde{\nu} = U_2(\theta)^\dagger \nu'$ and Hamiltonian $\tilde{H}$
\begin{equation}
\tilde{H} = \left(
\begin{array}{cc}
H & 0\\
0 & \Delta_{31}/2E + s_{13}^2 V
\end{array}
\right),
\end{equation}
where the block is given by
\begin{equation}
H = \frac{ 1 }{2E} 
\left(
\begin{array}{cc}
s_{12}^2\Delta_{21} +c_{13}^2 2EV& s_{12}c_{12}\Delta_{21}\\
s_{12}c_{12} \Delta_{21} & c^2_{12}\Delta_{21}
\end{array}
\right). \label{H_block}
\end{equation}
Through the definition of $\nu'$, we directly relate the propagation basis to the flavor basis via $\tilde{\nu} = \tilde U^\dagger  \nu$ where
\begin{eqnarray}
\tilde U &=& U_1(\theta_{23}) D_\delta U_2(\tilde \theta_{13}) \\
&=&  \left(  \begin{array}{ccc}
\tilde{c}_{13} & 0 & \tilde{s}_{13}\\
-\tilde{s}_{13} s_{23} e^{i\delta} & c_{23} & \tilde{c}_{13} s_{23} e^{i\delta}\\
-\tilde{s}_{13} c_{23} e^{i\delta} & -s_{23} & \tilde{c}_{13}c_{23} e^{i\delta}\\
\end{array}
\right) .
\end{eqnarray}
Thus, electron and muon neutrinos can be written in the local propagation basis as
\begin{equation}
\tilde{\nu}_{e} = \tilde{U}^\dagger \nu_e =  \left( \begin{array}{c}
\tilde{c}_{13} \\
0\\
\tilde{s}_{13}
\end{array} \right), \quad \tilde{\nu}_{\mu} = \tilde{U}^\dagger \nu_\mu =  \left( \begin{array}{c}
-\tilde{s}_{13} s_{23} e^{-i\delta} \\
c_{23} \\
\tilde{c}_{13} s_{23} e^{-i\delta}\end{array} \right)  \label{proj_nu} .
\end{equation}

As the correction $\theta$ depends upon the local density, we must consider
its
temporal (spatial) derivative in the evolution equation
\begin{equation}
\partial_t \tilde {\nu} = \partial_t [U_2(\theta)^\dagger] \nu'+
U_2(\theta)^\dagger \partial_t \nu'.
\end{equation}
Letting $\lambda_2$ be the generator of the rotation so that $U_2(\theta) =
e^{i \theta \lambda_2}$, we have
\begin{equation}
\partial_t U_2(\theta)^\dagger = -i \lambda_2 U_2(\theta)^\dagger \partial_t
\theta.
\end{equation}
Dropping insignificant terms, one may write the evolution equation in the
propagation basis as
\begin{equation}
i \partial_t \tilde {\nu} = (\tilde{H} + \lambda_2 \partial_t \theta) 
\tilde {\nu}.
\end{equation}

Considering only propagation through matter of constant density, 
the term $\partial_t \theta$ vanishes, and our evolution equation is
\begin{equation}
i \partial_t \tilde {\nu} = \tilde{H} \tilde {\nu}. \label{prop2}
\end{equation}
The block $H$ in this Hamiltonian can be easily diagonalized in closed form
with eigenvalues $\lambda_\pm$.  Of dynamical 
relevance is the difference in these
eigenvalues which yields the effective constant density mass-squared 
difference
\begin{equation}
\Delta_{21}^m = \Delta_{21} \sqrt{\cos^2 {2\theta_{12}}(1-E/E_R)^2 + 
\sin^2{2\theta_{12} }}, \label{d21m}
\end{equation}
where we have defined the resonance energy to be
\begin{equation}
E_R = \frac{\Delta_{21}\cos 2\theta_{12}}{2 V c_{13}^2}.  
\end{equation}
Fixing the solar mixing angle $\theta_{12} = 0.58$, we find the resonance 
energy
in the mantle of density $4.5$ g/cm$^3$ to be $E_R \sim 100$ MeV; in the core of density $\rho=11.5$ g/cm$^3$, the value is $E_R\sim 40$ MeV.
The mixing angle which achieves this diagonalization satisfies
\begin{equation}
\sin 2\theta_{12}^m = \frac{\sin 2 \theta_{12}}{\sqrt{\cos^2 2\theta_{12} (1-  
E/E_R)^2 + \sin^2 2 \theta_{12} }}. \label{th12m}
\end{equation}
At resonant energy, the  effective mixing angle in matter, $\theta_{12}^m$, results in maximal mixing for these two neutrino states in the propagation basis; this is termed the MSW resonance.
Additionally, matter effects require an accommodation to the
other mass-squared difference $\Delta_{31}^m = \Delta_{31}-2E\lambda_-$, though this correction is dominated by the vacuum value of the mass-squared difference.

In the analytic work that follows, we will be primarily interested in the
oscillatory region for the small mass-squared difference $\Delta_{21}$. We
will assume that the oscillations due to the two larger mass-squared
differences cannot be resolved at the baselines of interest $L$; that is, we will take
\begin{eqnarray}
\left\langle \sin^2\left({\frac{\Delta_{31}L}{4E}}\right) \right\rangle = \left\langle \sin^2\left({\frac{\Delta_{32}L}{4E}}\right)  \right\rangle
= \frac{1}{2}\\
\left\langle \sin \left({\frac{\Delta_{31}L}{4E}}\right) \right\rangle = \left\langle \sin \left({\frac{\Delta_{32}L}{4E}}\right)  \right\rangle= 0.
\end{eqnarray}
The upshot is that for sub-GeV neutrinos traveling through the earth the propagation basis provides us with a density-dependent effective two neutrino framework.

\section{Two flavor parametric resonance}

We will review parametric resonance within the context of a pure two neutrino system, say, $\nu_e$ and $\nu_\mu$, and briefly rehash known results.  This construction can be suitably adapted to describe parametric resonance in an effective two-neutrino framework for sub-GeV neutrinos traveling though matter of terrestrial densities.  We will denote the lone mixing angle as $\theta$ and mass-squared difference $\Delta$.    An exact solution for two neutrinos traveling through a castle wall potential was developed in Ref.~\cite{akhmedov_1} and expounded upon in Ref.~\cite{akh_long}.    

Following Ref.~\cite{floquet}, we will review the exact solution for neutrinos traversing a {\em general} periodic potential and then specify to the castle wall solution.   Without loss of generality, one may choose the neutrino Hamiltonian to be traceless.  If the two-neutrino Hamiltonian $H$ is not traceless from the start, one may add to the Hamiltonian with impunity any multiple of the identity, in particular $-\frac{1}{2} \mathrm{tr}(H) \mathbb{I}$; upon solving for the time evolution of the system, such multiples result  in an immeasurable overall phase.  
Thus, we may take the Hamiltonian to be of the form
\begin{equation}
H = \left( \begin{array}{cc} -\alpha(x) & \beta(x) \\ \beta(x) & \alpha(x) \end{array} \right), \label{H_can}
\end{equation}
where the real functions $\alpha(x)$ and $\beta(x)$ may depend on position by virtue of their density dependence.  If the Hamiltonian is expressed in the flavor basis, then these functions are
\begin{equation}
\alpha(x) = \frac{\Delta}{4E}\cos(2 \theta)-\frac{1}{2}V(x), \qquad \beta = \frac{\Delta}{4E}\sin(2 \theta).
\end{equation}
Denoting the period of the Hamiltonian as $L$, we have $H(x)=H(x+L)$.  

As the Hamiltonian is Hermitian, the evolution of the system is unitary $\nu(x) = \mathcal{U}(x)\nu(0)$.  Given this, we may use the Pauli matrices to write evolution through one period
\begin{equation}
\mathcal{U}(L) = Y - i \boldsymbol{\sigma}\cdot \bold{X};  \label{pauli_rep}
\end{equation}
 unitarity demands that the real quantities satisfy
\begin{equation}
Y^2 + | \bold{X}|^2 = 1.
\end{equation}
This can be written in terms of a phase $\Phi$
\begin{equation}
\mathcal{U}(L)= \exp\left[-i(\boldsymbol{\sigma} \cdot \hat{\bold X}) \Phi \right]
\end{equation}
with unit vector $\hat{\bf{X}} = \bold{X}/|\bold{X}|$ and
\begin{equation}
\cos\Phi = Y, \qquad \sin \Phi = |\bold{X}|. \label{phi_defn}
\end{equation}
In this last formulation, it is quite easy to see that, after $k$ periods, the evolution operator
can be written as
\begin{equation}
\mathcal{U}(kL)=[\mathcal{U}(L)]^k= \exp\left[-i(\boldsymbol{\sigma} \cdot \hat{ \bold X}) k \Phi \right]. \label{Phi_defn}
\end{equation}
Thus, if the neutrino state is initially $\nu (0)=\nu_e = (1,0)^T$, then after $k$ periods the state of the system is
\begin{equation}
\nu(kL) = \left( \begin{array}{c}
\cos(k\Phi)-i\hat{X}_3 \sin(k\Phi)\\
(\hat X_2- i \hat X_1) \sin(k\Phi)
\end{array}
\right).
\end{equation}
As $\hat {\bold{X}}$ is a unit vector by definition, we can redefine the terms which involve $\hat X_1$ and $\hat X_2$ by introducing a phase $\gamma$
\begin{equation}
\hat{X}_1 + i \hat{X}_2 = e^{i \gamma}\sqrt{1-\hat X_3^2} , \label{gamma_defn}
\end{equation}
so that we may rewrite the neutrino at baseline $kL$ as
\begin{equation}
\nu(kL) = \left( \begin{array}{c}
\cos(k\Phi)-i\hat{X}_3 \sin(k\Phi)\\
-ie^{i \gamma}\sqrt{1-\hat X_3^2}  \sin(k\Phi)
\end{array}
\right).
\end{equation}
In this form, it is clear that a maximum oscillation  $\nu_e \to \nu_\mu$ can be achieved if $\hat X_3 =0$.  This is the condition for parametric resonance for a general periodic Hamiltonian.

An analytical expression for $\hat X_3$ is hard to come by for a general density profile; however, a tractable solution does exist for the castle wall potential \cite{akhmedov_1,akh_long,floquet}.  Explicitly, the castle wall potential is defined as the periodic piecewise-constant function given by
\begin{equation}
V(x) = \left\{ \begin{array}{l} 
V_A \text{ for } 0 \le x < L_A \\ 
V_B \text{ for } L_A \le x < L_B \end{array} \right. \label{cwall}
\end{equation}
with the periodicity condition $V(x+L)=V(x)$ where  $L=L_A+L_B$.

Within one of the constant density regions, the effective mass-squared difference $\Delta_{A,B}$ and mixing angle $\theta_{A,B}$ in matter can be determined by diagonalizing the Hamiltonian as in Eqs.~(\ref{d21m},\ref{th12m}).  The evolution operator through one period is  composed of the constant density evolution operators $\mathcal{U}^{A,B}(x)$
\begin{equation}
\mathcal{U}(L) = \mathcal{U}^B(L_B) \mathcal{U}^A(L_A)  \label{one_per}
\end{equation}
where the constant density operators can be expressed as
\begin{equation}
\mathcal{U}^A(L_A) = \left(\begin{array}{cc}
c_A+i c_{2\theta_A} s_A & -i s_{2\theta_A} s_A\\
-i s_{2\theta_A} s_A& c_A - i c_{2\theta_A} s_A
\end{array}
  \right)\label{const_dens_ev}
\end{equation}
with the dynamic terms defined to be $c_A = \cos \varphi_A$ and $s_A = \sin \varphi_A$ where $\varphi_A = \Delta_A L_A/4E$.  An analogous expression exists for $\mathcal{U}^B(L_B)$.  Using the properties of Pauli matrices, one can express 
$\mathcal{U}(L)$ in the form of Eq.~(\ref{pauli_rep}) \cite{akhmedov_1,akh_long} with
 \begin{eqnarray}
 Y &=& c_A c_B - (s_{2\theta_A} s_{2\theta_B}+ c_{2\theta_A}c_{2\theta_B} )s_A s_B   \label{y_cw} \\
 \bold{X} &=& \left( \begin{array}{c}  
 s_A c_B s_{2\theta_A} + s_B c_A s_{2\theta_B}  \\
 (  s_{2\theta_B}  c_{2\theta_A}- s_{2\theta_A}  c_{2\theta_B}  ) s_A s_B \\
 -s_A c_B c_{2\theta_A}- s_B c_A c_{2\theta_B} 
 \end{array} \right). \label{xvec}
 \end{eqnarray}
 Thus, the condition for parametric resonance in a castle wall potential is
 \begin{equation}
 s_A c_B c_{2\theta_A}+ s_B c_A c_{2\theta_B} = 0.
 \end{equation}

\begin{figure}[th]
\includegraphics[width=8.6cm]{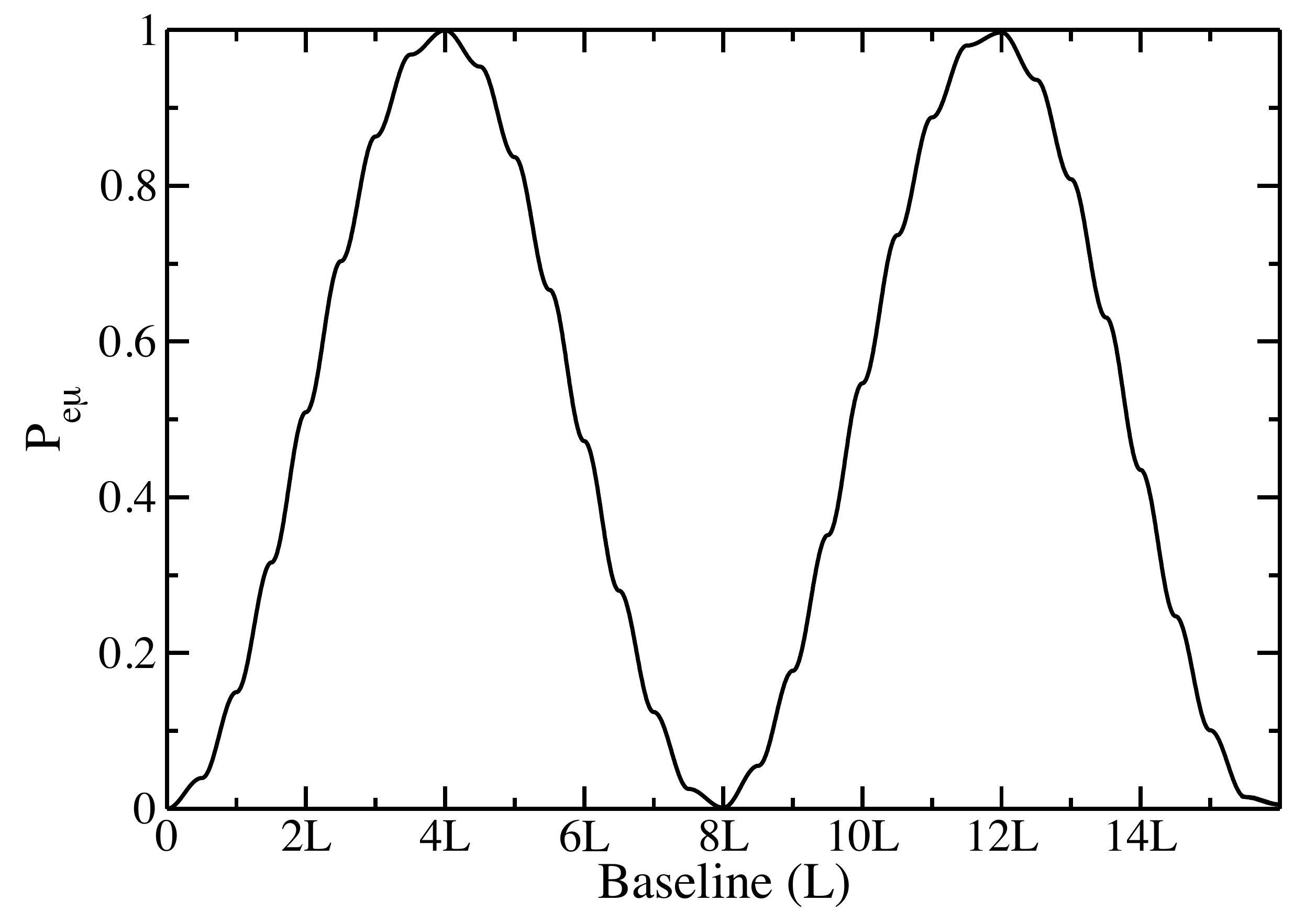}
\caption{Oscillation probability $\nu_e \to \nu_\mu$ through a castle wall potential using the following input:  $\theta=0.1$, $\Delta = 7.6\times 10^{-5}$ eV$^2$, $E = 200$ MeV, $\rho_A =0$ g/cm$^3$, $L_A =3255$ km, $\rho_B =10$ g/cm$^3$, $L_B =3212$ km. \label{halfwlfig} }
\end{figure}

Parametric resonance can be achieved via the ``half-wavelength condition" in which $L_A$ and $L_B$ are equal to an integer plus one-half (local) oscillation wavelengths; this amounts to $c_A=0=c_B$.  Such a half-wavelength scenario is pictured in Fig.~\ref{halfwlfig} with a vacuum mixing angle of $\theta=0.1$.  With this mixing angle, the maximum vacuum oscillation probability for $\nu_\mu$ appearance would be 0.04, yet after four periods of the castle wall potential, the  probability is unity.  In general, the oscillation does not attain unity at the end of a period but, rather, at some point in between.  For a profile satisfying the half-wavelength condition, one may determine from the definition of $\Phi$, Eq.~(\ref{phi_defn}), and the expression for $Y$, Eq.~(\ref{y_cw}), that the oscillation probability will be unity at the end of the $k$th period in the event that there exists an integer $n$ such that
\begin{equation}
2k | \theta_A -\theta_B|  = \left(n + \frac{1}{2} \right) \pi.
\end{equation}
For the parameters used to generate Fig.~\ref{halfwlfig}, it just so happens that the above is approximately satisfied for $k=4$, i.e., $8| \theta_A -\theta_B| = 3.49\, \pi $.

\begin{figure}[th]
\includegraphics[width=8.6cm]{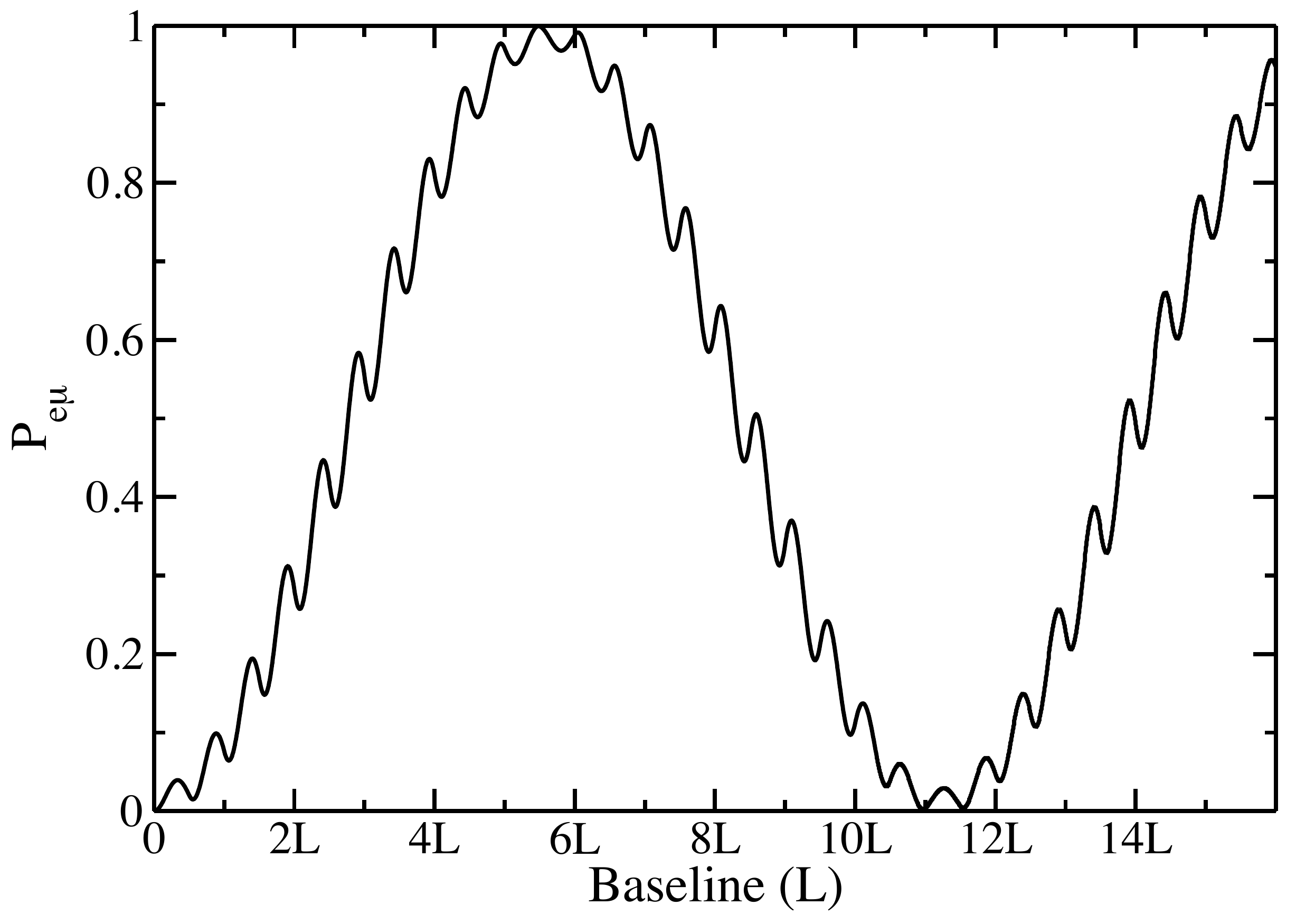}
\caption{Oscillation probability $\nu_e \to \nu_\mu$ through a castle wall potential using the following input:  $\theta=0.1$, $\Delta = 7.6\times 10^{-5}$ eV$^2$, $E = 200$ MeV, $\rho_A =0$ g/cm$^3$, $L_A =4882$ km, $\rho_B =10$ g/cm$^3$, $L_B =4819$ km. \label{genprfig} }
\end{figure}

More generally, the condition for parametric resonance  can be satisfied whenever
 \begin{equation}
\tan{\varphi_B}   = -\frac{c_{2\theta_A}}{c_{2\theta_B}} \tan{\varphi_A}.
 \end{equation}
 In Fig.~\ref{genprfig}, we demonstrate such a scenario.  Of note in this example is the fact that the oscillation probability reaches unity  roughly halfway between the fifth and sixth period of the matter potential.

\section{Three flavor parametric resonance}

Considering all three flavors, we can now study parametric resonance for sub-GeV neutrinos traveling through a castle wall potential for densities less than $15$ g/cm$^3$. 
We saw above that for sub-GeV neutrinos the relevant oscillations can be cast, to an good approximation, in the form of two-neutrino oscillations after rotating to the propagation basis.  As the rotation to the propagation basis is density dependent, we must use a transition matrix at the boundaries of the regions of constant density.
In our analytic treatment, we will make use of this and further approximations; however, our numerical computations will model a full three-neutrino system sans approximation.

We consider the same castle wall potential as in Eq.~(\ref{cwall}). 
 Using the sub-GeV approximation, the neutrino state after one period is given by
\begin{equation}
\nu(L) = \tilde{U}_B e^{-i \tilde{H}_B L_B}  U_2(\theta_B-\theta_A ) e^{-i \tilde{H}_A L_A}  \tilde{U}_A^\dagger \nu(0)
\end{equation}
where $\tilde{U}_{A,B}$ is evaluated for the constant potential $V_{A,B}$ and $\theta_{A,B}$ represent the matter corrections to $\theta_{13}$, Eq.~(\ref{theta}), for $V_{A,B}$.  To leading order in $\epsilon \theta_{13} $, the transition matrix between the two regions is
\begin{equation}
U_2(\theta_B-\theta_A ) \approx  \mathbb{I} + \frac{2\theta_{13} E(V_B-V_A)}{\Delta_{31}} \lambda_2 . \label{trans}
\end{equation}
The evolution operator in the propagation basis within a constant density region is
\begin{equation}
\mathcal{\tilde{U}}^A(x) := e^{-i \tilde{H}_A x} = \left(  
\begin{array}{ccc}
\mathcal{\tilde{U}}_{ee}(x) & \mathcal{\tilde{U}}_{e\mu}(x) & 0\\
\mathcal{\tilde{U}}_{\mu e}(x) & \mathcal{\tilde{U}}_{\mu \mu}(x) & 0\\
0 & 0 & \mathcal{\tilde{U}}_{\tau \tau}(x)
\end{array}
\right); \label{time_ev_prop}
\end{equation}
we make an analogous definition for $\mathcal{\tilde{U}}^B(x) = e^{-i \tilde{H}_B x}$. 
After one period,
the neutrino state is to leading order
\begin{equation}
\nu(L) = \tilde{U}_B   \mathcal{\tilde{U}}(L)   \tilde{U}_A^\dagger \nu(0)  +  \mathcal{O}(\theta_{13} \epsilon), \label{lot}
\end{equation}
where the leading order contribution to the evolution operator is denoted by $\mathcal{\tilde{U}}(L) = \mathcal{\tilde{U}}^B(L_B) \mathcal{\tilde{U}}^A(L_A)$, consistent with Eq.~(\ref{one_per}).
This is the dominant contribution to sub-GeV oscillations in the earth, but we examine the $\mathcal{O}(\epsilon \theta_{13})$ correction.
Returning to the transition matrix, Eq.~(\ref{trans}), we note
\begin{equation}
 \tilde{\mathcal{U}}^B(L_B)  \lambda_2   \tilde{\mathcal{U}}^A(L_A)  = \left( \begin{array}{ccc}
0 & 0 &  \tilde{\mathcal{U}}^B_{ee}  \tilde{\mathcal{U}}^A_{\tau \tau} \\
0 & 0 &  \tilde{\mathcal{U}}^B_{ \mu e}  \tilde{\mathcal{U}}^A_{\tau \tau} \\
-  \tilde{\mathcal{U}}^B_{\tau \tau}  \tilde{\mathcal{U}}^A_{ee}  & -  \tilde{\mathcal{U}}^B_{\tau \tau}  \tilde{\mathcal{U}}^A_{e\mu} & 0 \\
\end{array}
\right)
\end{equation}
where $\mathcal{\tilde{U}}^A_{\alpha \beta} =\mathcal{\tilde{U}}_{\alpha \beta}^A(L_A)$, etc.
Supposing $\nu(0)=\nu_e$, the probability of detecting $\nu_\mu$ is 
\begin{eqnarray}
\mathcal{P}_{e\mu}(L) &=& | \tilde{s}_{13_B} s_{23} \tilde{c}_{13_A}e^{i\delta} \tilde{\mathcal{U}}_{ee}(L)  - c_{23} \tilde{c}_{13_A} \tilde{\mathcal{U}}_{\mu e}(L)  |^2 \nonumber \\
&&+ \tilde{s}_{13_A}^2 \tilde{c}_{13_B}^2 s_{23}^2  \nonumber \\
&& + (\theta_B - \theta_A)^2 \tilde{c}_{13_B}^2 s_{23}^2 \tilde{c}_{13_A}^2 | \tilde{\mathcal{U}}_{ee} (L)|^2, \label{pem_ax1}
\end{eqnarray}
assuming the ``atmospheric" oscillations  average to zero $\langle \tilde{\mathcal{U}}_{\tau \tau} \rangle\sim 0$ and neglecting terms $\mathcal{O}(\theta_{13}^4)$.
The term in the oscillation probability proportional to $(\theta_B - \theta_A)^2$ is the leading order remnant from the transition matrix between the constant density regions, and its size is order $\mathcal{O}(\theta_{13}^2 \epsilon^2)$. Neglecting this term will result in a discontinuity in the oscillation probability across the boundary between regions; however, the $\mathcal{O}(\epsilon^2)$ factor suppresses the significance of this discontinuity.  We shall neglect this term in our semi-analytic analysis; this amounts to setting the transition matrix to the identity.  

With this simplification, the neutrino system in the propagation basis effectively consists of two states as the evolution of the  $\tilde{\nu}_\tau$ state decouples.  The analysis of two-neutrino parametric resonance can be carried over wholesale with one adjustment for the effective potential in the propagation basis, namely, $V_{A,B} \mapsto \tilde{c}^2_{{13}_{A,B}} V_{A,B}$.   Given this, if one begins with an initial state $\nu(0)=\nu_e$, then the neutrino state in the propagation basis at $x=kL$ is
\begin{equation}
\tilde{\nu}(kL)  = \left( \begin{array}{c}
\tilde c_{{13}_A}[\cos(k\Phi)-i\hat{X}_3 \sin(k\Phi)]\\
\tilde c_{{13}_A}[-ie^{i \gamma}\sqrt{1-\hat X_3^2}  \sin(k\Phi)]\\
\tilde s_{{13}_A} \tilde{\mathcal{U}}_{\tau \tau}(kL)
\end{array}
\right) . \label{state_kl}
\end{equation}
The probability for a $\nu_\mu$ detection after $k$ periods is  
\begin{eqnarray}
\mathcal{P}_{e \mu}(kL) &=& \tilde c_{{13}_A}^2  \Big{|}  \tilde{s}_{{13}_B} s_{23} [\cos(k\Phi)-i\hat{X}_3 \sin(k\Phi)]  \nonumber \\
&&+ ie^{i (\gamma-\delta)}c_{23}  \sqrt{1-\hat X_3^2}  \sin(k\Phi) \Big{|}^2   \nonumber \\
&&+ \tilde{c}_{{13}_B}^2 \tilde s_{{13}_A}^2 s^2_{23}.
 \end{eqnarray}

This expression for the oscillation probability is considerably more complicated than its purely two-neutrino analogue.  In particular, for arbitrary vacuum mixing angles and mass-squared differences, it is clear that the appearance oscillation probability $\nu_e \to \nu_\mu$  cannot generally become unity via parametric resonance.  This is no surprise, given the additional $\nu_\tau$ oscillation channel.  Still, the question remains as to how a particular matter profile might maximally enhance the oscillation through parametric resonance.  To develop a parametric resonance condition, it is best to examine the oscillation probability to leading order in $\theta_{13}$
\begin{widetext}
\begin{equation}
\mathcal{P}_{e \mu}(kL) = \sin^2(k\Phi) c_{23}^2 (1-\hat{X}_3^2)\left[ 1-\frac{2 \theta_{13} s_{23} \cos(\gamma-\delta)\hat{X}_3}{c_{23} \sqrt{1-\hat{X}_3^2}} \right]  -2 \sin(k\Phi) \cos(k\Phi) \theta_{13} s_{23}c_{23}  \sin(\gamma-\delta) \sqrt{1-\hat X_{3}^2} 
+\mathcal{O}(\theta_{13}^2) . \label{pem3}
 \end{equation}
 \end{widetext}
Overall, all terms in Eq.~(\ref{pem3}) are modulated by (at least) one factor of $\sqrt{1-\hat X_3^2}$, so that the condition for parametric resonance will be a perturbation of the purely two-neutrino condition, $\hat X_3=0$.   In fact, one can show that the value of $\hat X_3$ which maximizes $\mathcal{P}_{e\mu}$ is of order $\mathcal{O}(\theta_{13})$.  The term in Eq.~(\ref{pem3}) which is linear in $\hat X_3$ also has an explicit factor of $\theta_{13}$. As our approximation is only valid up to $\mathcal{O}(\theta_{13})$, we must, for consistency's sake, effectively adopt the purely two-neutrino condition for parametric resonance, $\hat X_{3}=0$.
Combining the remaining terms, we find, consistent to our level of approximation, the oscillation probability to be
\begin{equation}
\mathcal{P}_{e \mu}(kL) = c_{23}^2 \sin^2[k\Phi + \psi]  \label{pem_3nu}
 \end{equation}
with $\psi =- \theta_{13}  \tan \theta_{23} \sin(\gamma-\delta)$ when $\hat{X}_3=0$.  Near the boundary of an integer number of periods, we see that the oscillation probability is bounded by $c_{23}^2$ rather than the unit bound in the purely two-neutrino framework.  Also, terms linear in $\theta_{13}$ which are CP odd, enter only as a phase shift in the oscillation probability.  

In this three neutrino system, the bound of $c_{23}^2$ in Eq.~(\ref{pem_3nu}) arises from the projection to the propagation basis which permitted the use of the two-neutrino analysis.   If we work within the effective two-neutrino picture in the propagation basis, then the amplitude of the oscillation probability is $\sin^2 2\theta_{12}$; as with the two neutrino case, parametric resonance allows one to  saturate the transition probability at unity.  Returning to the flavor basis, we then expect the probability to saturate at $c_{23}^2$.
It is useful to compare Eq.~(\ref{pem_3nu}) with the oscillation probability for sub-GeV neutrinos in matter of constant density.
Referring to Ref.~\cite{dcl_onechannel}, let us only consider  the $\nu_e \to \nu_\mu$ oscillation probability for the situation in which CP is maximally violated with $\delta=\frac{\pi}{2}$; for other values of $\delta$, similar arguments hold.
To leading order in $\theta_{13}$, we have 
\begin{equation}
\mathcal{P}_{e\mu_{\delta=\frac{\pi}{2}}}  \approx  \sin^2 2\theta_{12} c_{23}^2\sin^2\left( \frac{\Delta^m_{21}L}{4E} + \phi\right) 
\end{equation}
with the phase $\phi\approx  - \theta_{13} \tan \theta_{23}/ \sin 2\theta_{12} $.  Via parametric resonance we can saturate the $\sin^2 2\theta_{12}$ bound; sending $\sin^2 2\theta_{12}\mapsto 1$, we find an expression similar to Eq.~(\ref{pem_3nu}).

\begin{figure}[th]
\includegraphics[width=8.6cm]{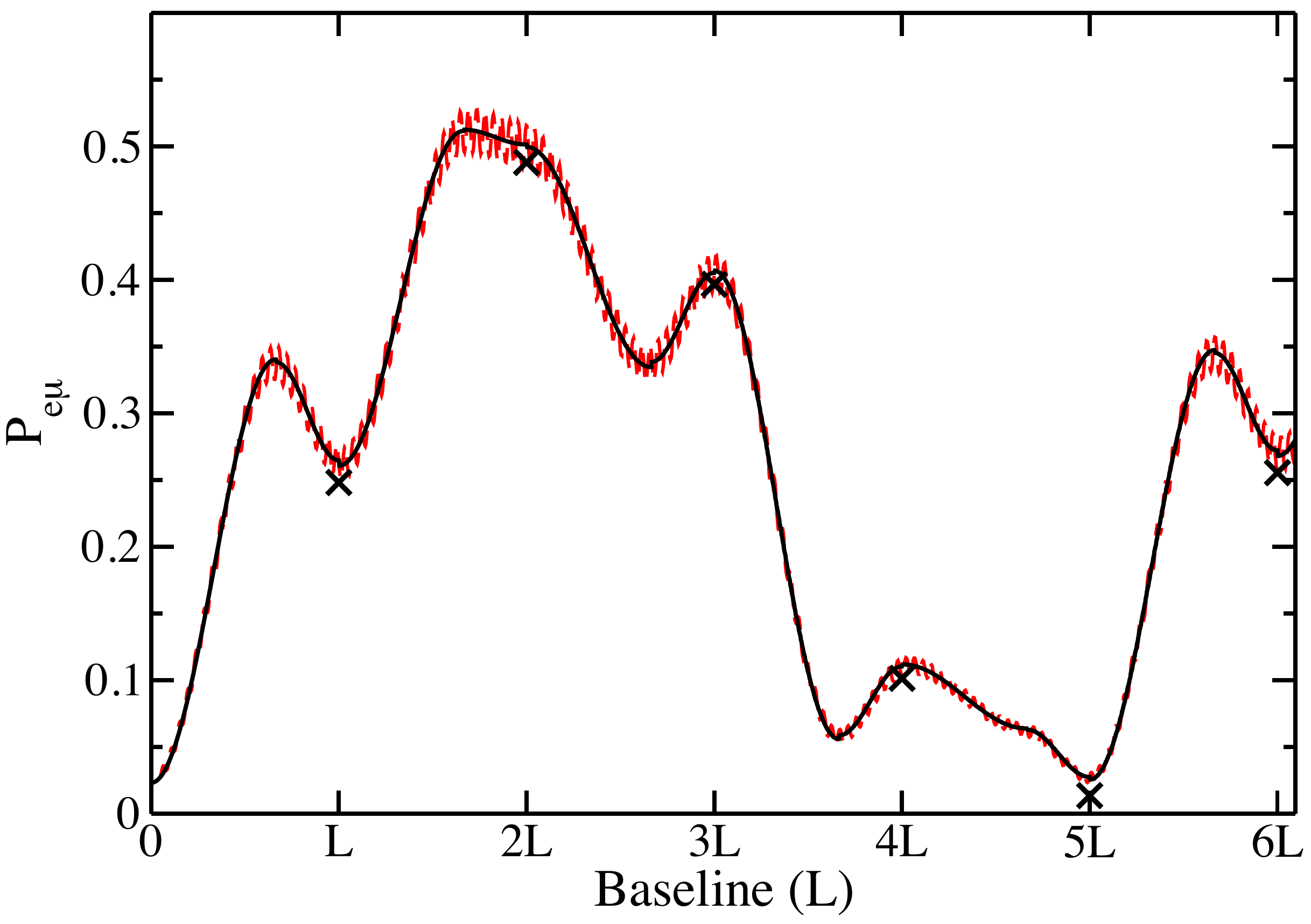}
\caption{(Color online) Oscillation probability $\nu_e \to \nu_\mu$ through a castle wall potential using the following  parameters:  $\theta_{12}=0.58$, $\theta_{13}=0.15$, $\theta_{23}=0.785$, $\delta=0$, $\Delta_{21} = 7.6\times 10^{-5}$ eV$^2$,  $\Delta_{31} = 2.4\times 10^{-3}$ eV$^2$, $E = 200$ MeV, $\rho_A =4.5$ g/cm$^3$, $L_A =3161$ km, $\rho_B =11.5$ g/cm$^3$, $L_B = 1597$ km.  The dashed [red] curve uses the exact three-neutrino framework, averaging over the $\Delta_{31}$ oscillations. The solid [black] curve employs the effective two-neutrino approximation in which the transition matrix between regions $A$ and $B$ is taken to be the identity.  The cross, $\boldsymbol{\times}$, represents the value of Eq.~(\ref{pem_3nu}) at points $kL$.
 \label{fig3}}
\end{figure}

In Fig.~\ref{fig3}, we compare the approximate analytic treatment for neutrinos traveling through a castle wall potential with exact numerical results. Realistic values of the mixing angles, mass-squared differences, and densities have been chosen so as to satisfy the half wavelength condition.  The parameters which are germane to our approximations have the values $\theta_{13}=0.15$ and $\epsilon:=2EV_B/\Delta_{31}=0.07$.  In the figure, we plot as the solid [black] curve the effective two-neutrino approximation in which we take as the identity the transition matrix between boundary layers.   This curve is superimposed upon the results of an exact three-neutrino computation, plotted as the dashed [red] curve.  For the three-neutrino curve, we average over the $\Delta_{31}$ oscillations so as to mimic a detector's finite energy resolution; remnants of these oscillations appear as the higher frequency wiggles in the curve.  The effective two-neutrino approximation accurately captures the oscillations driven by the $\Delta_{21}$ mass-squared difference.  There is a slight discontinuity in the solid curve at the boundary between regions $A$ and $B$; however, it is not too severe as the product $\epsilon \theta_{13}$ is rather small.  Finally, keeping only terms linear in $\theta_{13}$, we were able to determine the oscillation probability after an integer number of periods, Eq.~(\ref{pem_3nu}).  In the figure, we plot the value of Eq.~(\ref{pem_3nu}) using the cross, $\boldsymbol{\times}$.  This agrees rather well with the other two curves though it does have a systematically lower value than the more exact treatments; we trace this to a positive term of order $\mathcal{O}(\theta_{13}^2)$ that has been neglected.

\begin{figure}[th]
\includegraphics[width=8.6cm]{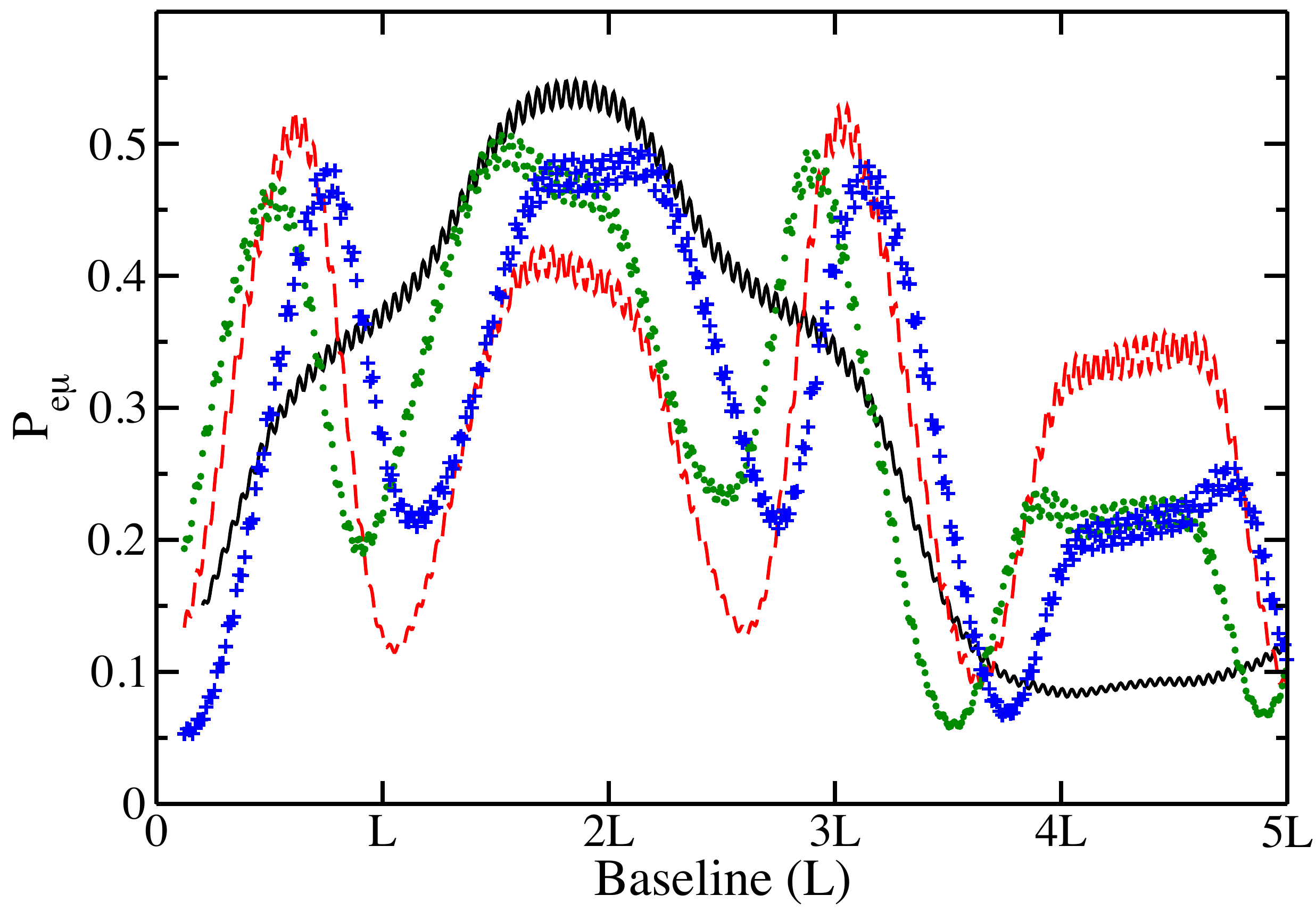}
\caption{(Color online) Oscillation probability $\nu_e \to \nu_\mu$ through a castle wall potential using the following common input:  $\theta_{12}=0.58$, $\theta_{13}=0.3$, $\theta_{23}=0.785$, $\Delta_{21} = 7.6\times 10^{-5}$ eV$^2$,  $\Delta_{31} = 2.4\times 10^{-3}$ eV$^2$, $E = 200$ MeV, $\rho_A =4.5$ g/cm$^3$, $L_A =3239$ km, $\rho_B =11.5$ g/cm$^3$, $L_B =1707$ km.  The solid [black] curve has $\delta=0$; the dashed [red] curve has $\delta=\pi$; the dotted [green] curve has $\delta=\frac{\pi}{2}$; and the $+$ [blue] curve has $\delta = \frac{3\pi}{2}$. The $\Delta_{31}$ oscillations are averaged over. \label{fig4}}
\end{figure}

In Fig.~\ref{fig4}, we implement parametric resonance, $\hat X_3=0$, via the half-wavelength condition, a three-neutrino analog to Fig.~\ref{halfwlfig}. For the neutrino oscillation parameters, we use the best fit values from the global analysis of oscillation data in Ref.~\cite{fogli_2011}, save $\theta_{13}$.  In order to accentuate the effects of this mixing angle, we set $\theta_{13}=0.3$ which is roughly 10-$\sigma$ larger than its best fit value \cite{fogli_2011}.  For this matter profile, the oscillation phase for a single period is $\Phi=2.49$, and the half-wavelength condition forces $\hat X_2 = 1$ so that $\gamma = \frac{\pi}{2}$.  With maximal mixing for $\theta_{23}$, the phase offset for the oscillations is $\psi = -0.3 \cos \delta$.  Despite the fact that this is relatively small, $|\psi| \le 0.3$, the phase can have a large impact as to where the (absolute) maximum oscillation probability occurs, since Eq.~(\ref{pem_3nu}) is a function of a discrete number of periods $kL$.  For four different values of $\delta$, we see widely varied traces for $\mathcal{P}_{e\mu}$.  The solid [black] curve plots the oscillation probability for the CP conserving case of $\delta =0$.  Focusing only upon the curves at points $kL$, this curve attains is maximum value near $x\approx 2L$, as $2\Phi+\psi \approx 1.5 \pi$.   As a contrast, the dashed [red] curve also plots CP conserving oscillations, but with $\delta = \pi$.  Relative to the first case, the sign of $\psi$ changes, and we find the maximum value of $\mathcal{P}_{e\mu}$ is attained for $x\approx 3L$ given that $3\Phi + \psi \approx 2.5 \pi$.   We also plot two cases of maximal CP violation.  The dotted [green] curve has $\delta = \frac{\pi}{2}$, and the $+$ [blue] curve has $\delta= \frac{3\pi}{2}$.   As $\psi =0$ for both of these cases,  they both intersect for integer multiples of the period.  Also, at these points, the oscillation probability takes a value intermediate of the two CP conserving cases.

More notable, perhaps, is the separation between the CP violating and CP conserving curves at the end of each period. For a general matter profile,  the difference between the CP conserving curves, i.e., $\delta =0$ and $\delta = \pi$ [or equivalently $\pm \theta_{13}$] at the point $kL$ is
\begin{equation}
\mathcal{P}_{e\mu_{\delta = 0}} - \mathcal{P}_{e\mu_{\delta = \pi}} \approx -\theta_{13} \sin 2 \theta_{23} \sin \gamma \sin 2k\Phi .
\end{equation}
The maximum separation between the two curves occurs when $k\Phi = n \frac{\pi}{2} + \frac{\pi}{4}$ for some integer $n$ and $\gamma = \frac{\pi}{2}, \frac{3\pi}{2}$
\begin{equation}
\left|\mathcal{P}_{e\mu_{\delta = 0}}  - \mathcal{P}_{e\mu_{\delta=\pi}}\right| \le |\theta_{13} \sin 2 \theta_{23}|.
\end{equation}
These conditions can be trivially satisfied with the half-wavelength condition.
Incidentally, for sub-GeV neutrinos propagating through a constant density region in the earth's mantle or core, it was shown in Ref.~\cite{latimer_th13th23} that
\begin{equation}
\left|\mathcal{P}_{e\mu_{\delta = 0}}  - \mathcal{P}_{e\mu_{\delta=\pi}}\right| \le |\theta_{13}  \sin{2\theta_{23}}  \sin{4\theta_{12}^m} |.
\end{equation}
For a mantle density of $\rho = 4.5$ g/cm$^3$, the difference between the oscillation probabilities in constant density matter for $\pm \theta_{13}$ is suppressed by a factor of $|\sin 4\theta_{12}^m| = 0.75$ relative to the castle wall profile, whereas for a density of $11.5$ g/cm$^3$, the suppression is a factor of 0.84.  

\begin{figure}[th]
\includegraphics[width=8.6cm]{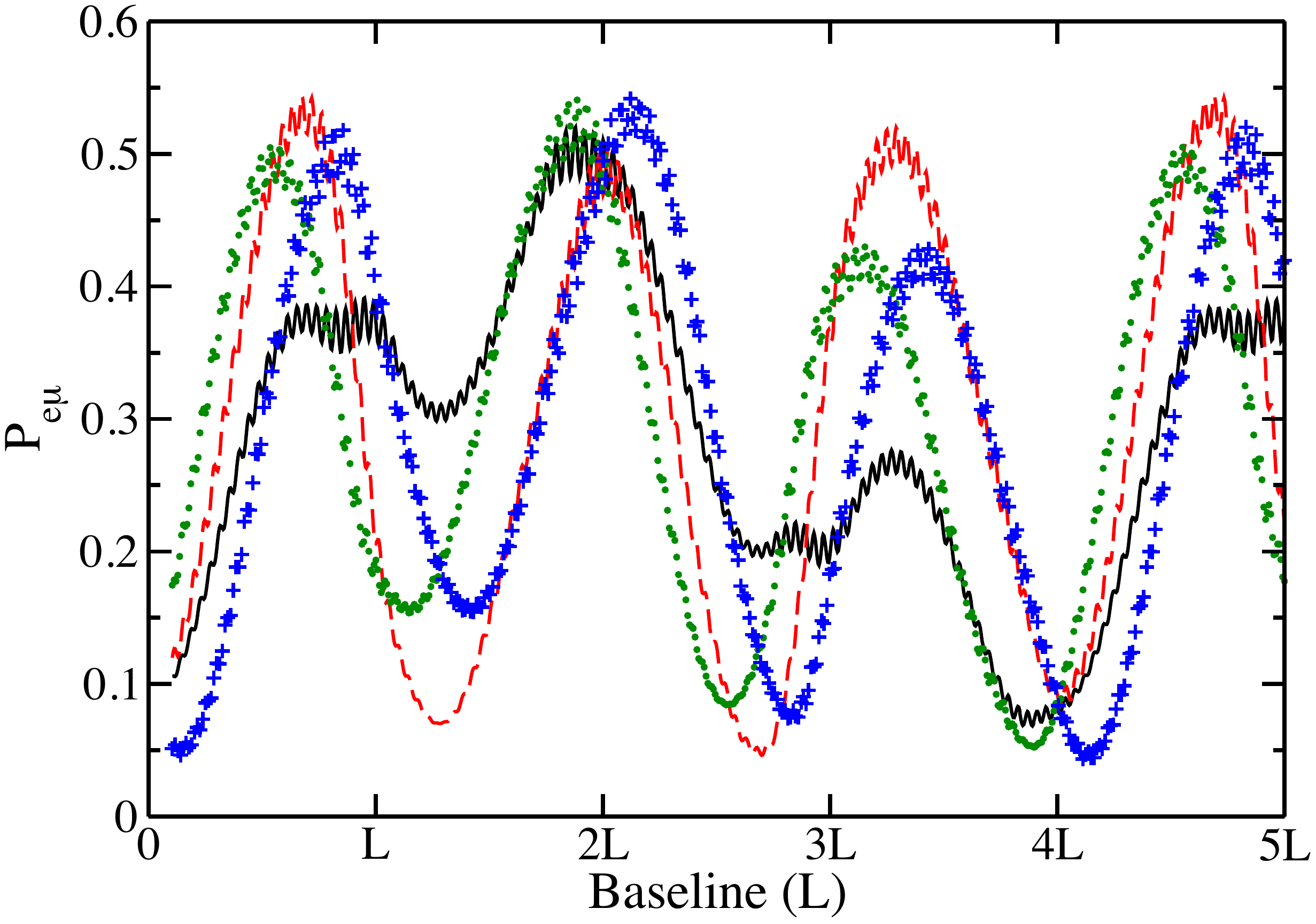}
\caption{(Color online) Oscillation probability $\nu_e \to \nu_\mu$ through a castle wall potential using the same oscillation parameters as in Fig.~\ref{fig4} but with $E = 200$ MeV, $\rho_A =3.8$ g/cm$^3$, $L_A =4024$ km, $\rho_B =11.5$ g/cm$^3$, $L_B =924$ km.  The curves are as in Fig.~\ref{fig4}.
 \label{fig5}}
\end{figure}

The separation between the maximal CP violating case $\delta=\frac{\pi}{2}$ and the CP conserving case $\delta=0$ at the end of the $k$th period is
\begin{equation}
\mathcal{P}_{e\mu_{\delta = 0}}  - \mathcal{P}_{e\mu_{\delta=\frac{\pi}{2}}}   \approx -\frac{1}{2} \theta_{13} \sin 2 \theta_{23} (\sin \gamma +\cos \gamma) \sin 2k\Phi.
\end{equation}
The maximum separation between these two curves occurs whenever $k\Phi = \frac{n\pi}{2} + \frac{\pi}{4}$ and $\gamma =\frac{\pi}{4},\frac{5\pi}{4}$; this separation is 
\begin{equation}
\Big|\mathcal{P}_{e\mu_{\delta = 0}}  - \mathcal{P}_{e\mu_{\delta=\frac{\pi}{2}}}\Big| \le \frac{1}{\sqrt{2}} |\theta_{13}  \sin{2\theta_{23}} |.
\end{equation}
Unlike the separation between the $\delta=0,\pi$ curves, the implementation of these constraints is nontrivial. For $k=1$, the requirement for $\Phi$ implies $\frac{1}{2}=\sin^2\Phi  =|\mathbf{X}|^2$.  With $\hat{X}_3=0$ and the requirement on $\gamma$, these constraints translate into $X_1^2 = \frac{1}{4} = X_2^2$.    For fixed vacuum values of the mixing angle $\theta_{12}$ and mass-squared difference $\Delta_{21}$, there are five remaining free parameters for a general castle wall profile:  the neutrino energy and the density and length of the two regions in the castle wall.  With the three constraints on the values of $X_j$, this leaves two free parameters, say, the two densities; however, for general values of the two densities, applying these constraints can result in a complex value for the neutrino energy.  Depending on the value of $\theta_{12}$, the densities $\rho_A$ and $\rho_B$ must be sufficiently different in order for the constraints to result in a real value for $E$.    In Fig.~\ref{fig5}, we show an example of these constraints which maximizes the separation between the $\delta=0$ [black] solid and $\delta=\frac{\pi}{2}$ [green] dotted curves at the end of each odd period.  The parameters which we employ to produce these curves results in the phase values $\Phi=3\pi/4$ and $\gamma=\pi/4$. If one knew the values of all the neutrino oscillation parameters, save the CP phase, this castle wall would could resolve $\delta=0$ and $\delta=\frac{\pi}{2}$, but a degeneracy in the parameter space would remain if measurements were made only at the boundaries because the $\delta=0,\frac{3\pi}{2}$ curves intersect as do the $\delta=\frac{\pi}{2},\pi$ curves.  

\begin{figure}[th]
\includegraphics[width=8.6cm]{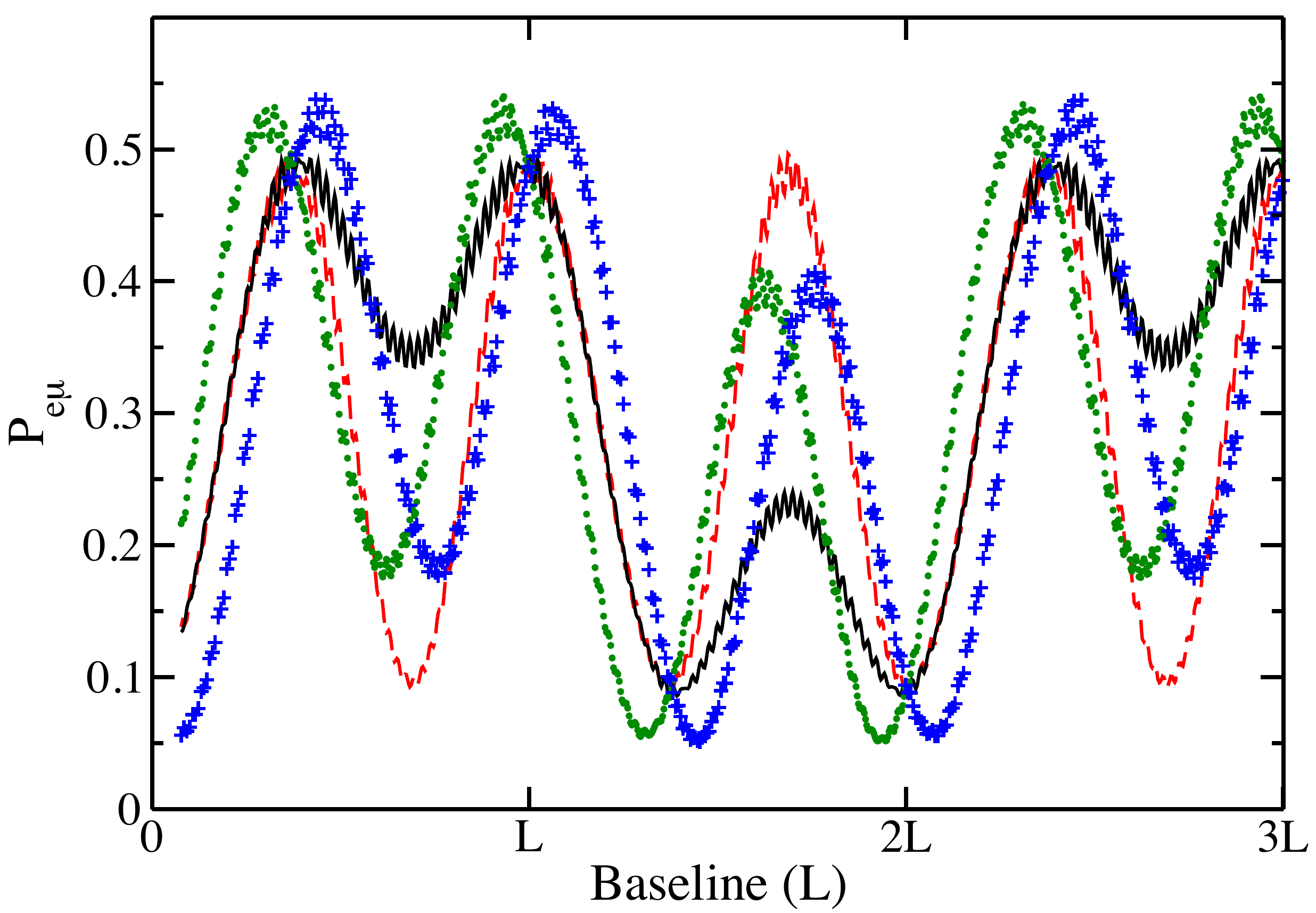}
\caption{(Color online) Oscillation probability $\nu_e \to \nu_\mu$ through a castle wall potential using the same parameters as Fig.~\ref{fig4} but with $E=98.5$ MeV, $L_A =1749$ km, and $L_B=2896$ km. \label{fig6}}
\end{figure}
As an aside, we  comment upon one additional case of parametric resonance which carries the phase condition $\gamma=0,\pi$; as in the previous example, this extra condition places severe restrictions on the baseline and neutrino energies.  From the definition of the phase $\gamma$ in Eq.~(\ref{gamma_defn}), it is clear that $\hat X_2$ must vanish when $\gamma=0,\pi$.  Thus, along with the condition for parametric resonance, one requires $\hat X_1 = 1$.  To satisfy all of these requirements, in one region, say region $A$, the width of the region $L_A$ must be an integer number of wavelengths so that $s_A =0$.  In the other region, the width of the region $L_B$ must be an integer-plus-one-half wavelengths so that $c_B=0$, {\em and} the energy of the neutrinos must be at the MSW resonance in region $B$ so that $c_{2\theta_B}=0$. In spirit, these criteria do not reflect true parametric resonance; rather this is essentially a manifestation of the MSW resonance.  Neutrinos which travel an integer number of wavelengths through region $A$ will exit the region in essentially the same state as they entered the region; then, in region $B$, we have explicitly required the energy to be at the MSW resonant value.  
Turning to $\Phi$, the conditions require  $\Phi=\frac{\pi}{2},\frac{3\pi}{2}$ so that
 it is not possible to simultaneously satifsy $\gamma=0,\pi$ and $k\Phi = \frac{m\pi}{2} + \frac{\pi}{4}$; thus, the curves for all values of $\delta$ will intersect at the boundary between periods. 
 In Fig.~\ref{fig6}, we implement this scenario. The neutrino energy is chosen to match the MSW resonance of region $A$, and $L_A$ is one-half wavelength in size.  For the next region, $L_B$ is chosen to be a full wavelength in size.  For the different values of the CP phase, we see that the oscillation probability is the same value, but that the curves are quite distinguishable in the interior of region $B$.

\section{Discussion}

For neutrino energies on the order of $\lesssim 1$~GeV, the  wavelength of vacuum neutrino oscillations associated with the mass-squared difference $\Delta_{21}$ is $ 3\times10^4$ km.  Given this scale, a closed laboratory demonstration of parametric resonance is not feasible.  The only recourse is to use the Earth's mantle-core-mantle transition to induce the resonance. Unfortunately, this means that the neutrinos can only travel through fewer than two periods of a castle wall potential; however, sub-GeV atmospheric neutrinos can still undergo significant enhancement due to parametric resonance \cite{liu_smirn, liu_mikh_smirn,petcov_param,akh_long}.
 
 \begin{figure}[th]
\includegraphics[width=8.6cm]{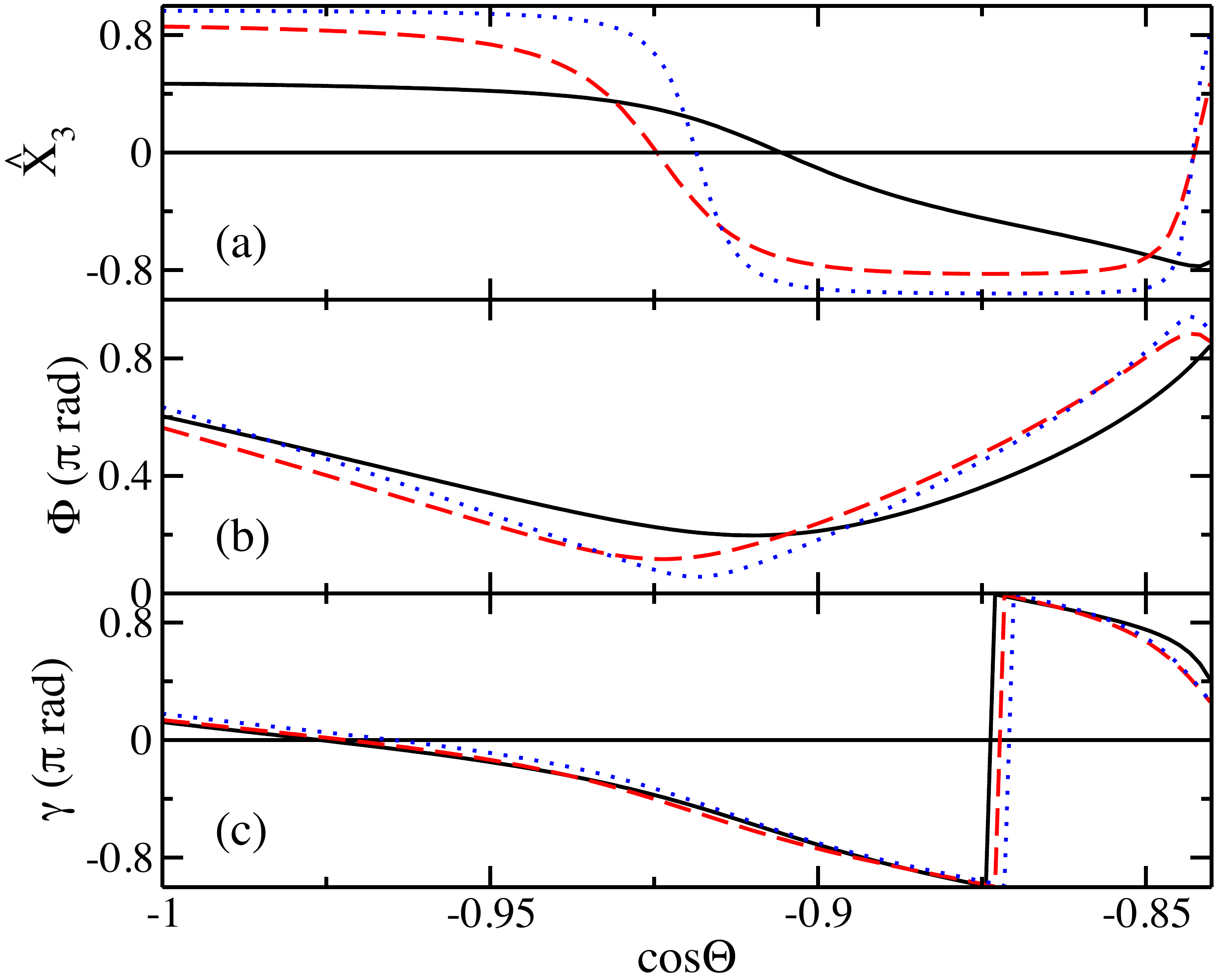}
\caption{(Color online) We plot (a) $\hat{X}_3$, (b) $\Phi$, and (c) $\gamma$ versus the cosine of the zenith angle for neutrinos traveling along a chord through the earth. For the solid [black] curves, the neutrino energy is 200 MeV; for the dashed [red] curves, 400 MeV; for the dotted [blue] curves, 800 MeV.   
 \label{fig7}}
\end{figure}

We model the earth as a  constant density core of radius $R_c=3485$ km and density $\rho_c=11.5$ g/cm$^3$ surrounded by a constant density mantle with $\rho_m=4.5$ g/cm$^3$.  The path of a neutrino, a chord through the earth's interior, can be parametrized via the zenith angle $\Theta$.  Neutrinos which traverse the entire diameter of the earth would have $\Theta=\pi$ and thus $\cos \Theta = -1$. For this trajectory,  the amount of the mantle seen by the neutrino before entering the core is $L_m = R_e - R_c = 2886$ km where the radius of the earth is $R_e = 6371$ km, and the neutrino traverses a path through the core of length $L_c = 2 R_c =6970$ km.  As $\Theta$ decreases, the mantle path length increases while the core path length decreases; generally, one has
\begin{eqnarray}
 L_m&=&-R_e \cos\Theta-\sqrt {R_c^2-(R_e \sin\Theta)^2},	\\
  L_c&=&2\sqrt {R_c^2-(R_e\sin\Theta)^2}.
\end{eqnarray}
At the zenith angle $\Theta_\text{crit}$, the neutrino's trajectory is tangential to the core; this occurs when $\sin \Theta_\text{crit} = R_c/R_e$; i.e., $\Theta_\text{crit} = 2.56$, or $\cos \Theta_\text{crit} = -0.84$.

\begin{figure}[th]
\includegraphics[width=8.6cm]{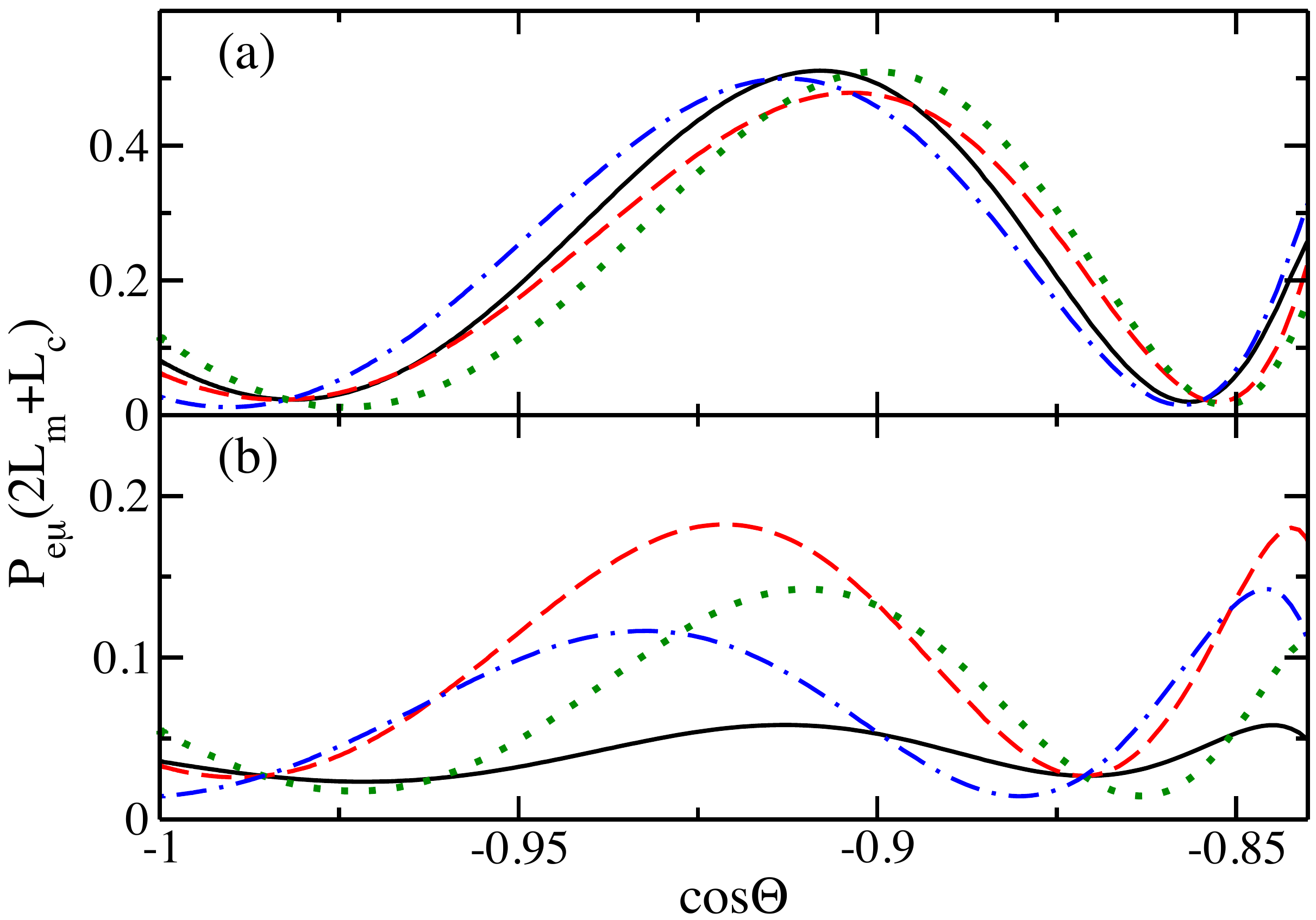}
\caption{(Color online) The oscillation probability $\mathcal{P}_{e\mu}$ for a neutrino which has traveled along a chord through the earth with zenith angle $\Theta$ for energies (a) 200 MeV and (b) 800 MeV. The solid [black] curves have $\delta=0$; the dashed [red] curves have $\delta=\pi$; the dotted [green] curves have $\delta=\frac{\pi}{2}$; the dot-dashed [blue] curves have $\delta = \frac{3\pi}{2}$. \label{fig8} }
\end{figure}

From these path lengths, we can compute the relevant data for parametric resonance as a function of $\Theta$.  In Fig.~\ref{fig7}, we plot $\hat{X}_3$, $\Phi$, and $\gamma$, respectively, for select neutrino energies from 200 MeV to 800 MeV.    From Fig.~\ref{fig7}(a), we see that the condition for parametric resonance is satisfied in this energy range for chords with zenith angles such that $-0.93 \le \cos \Theta \le -0.9$;  this is consistent with the neutrino oscillograms in Refs.~\cite{akh_12,akh_nu_osc_cpv}.  The parametric resonance is reflected in the enhancement of the $\nu_e \to \nu_\mu$ oscillation probability that would be seen at a detector located at the end of the chord after the neutrino has traveled a total distance of $2L_m+L_c$.  In Fig.~\ref{fig8}, this enhancement is apparent as we plot, as a function of $\cos \Theta$, the detector value of $\mathcal{P}_{e\mu}$ for 200 MeV and 800 MeV neutrinos for various values of the CP phase $\delta$.  The average location of the peak values of  $\mathcal{P}_{e\mu}$ for the different values of $\delta$ corresponds to the zenith angle at which $\hat X_3=0$.  Though the overall amplitude of the 800 MeV curves are suppressed  because the energy is far from the MSW resonance, it is noteworthy that there is a large separation between the peak values of the oscillation probability for different values of $\delta$.

\begin{figure}[th]
\includegraphics[width=8.6cm]{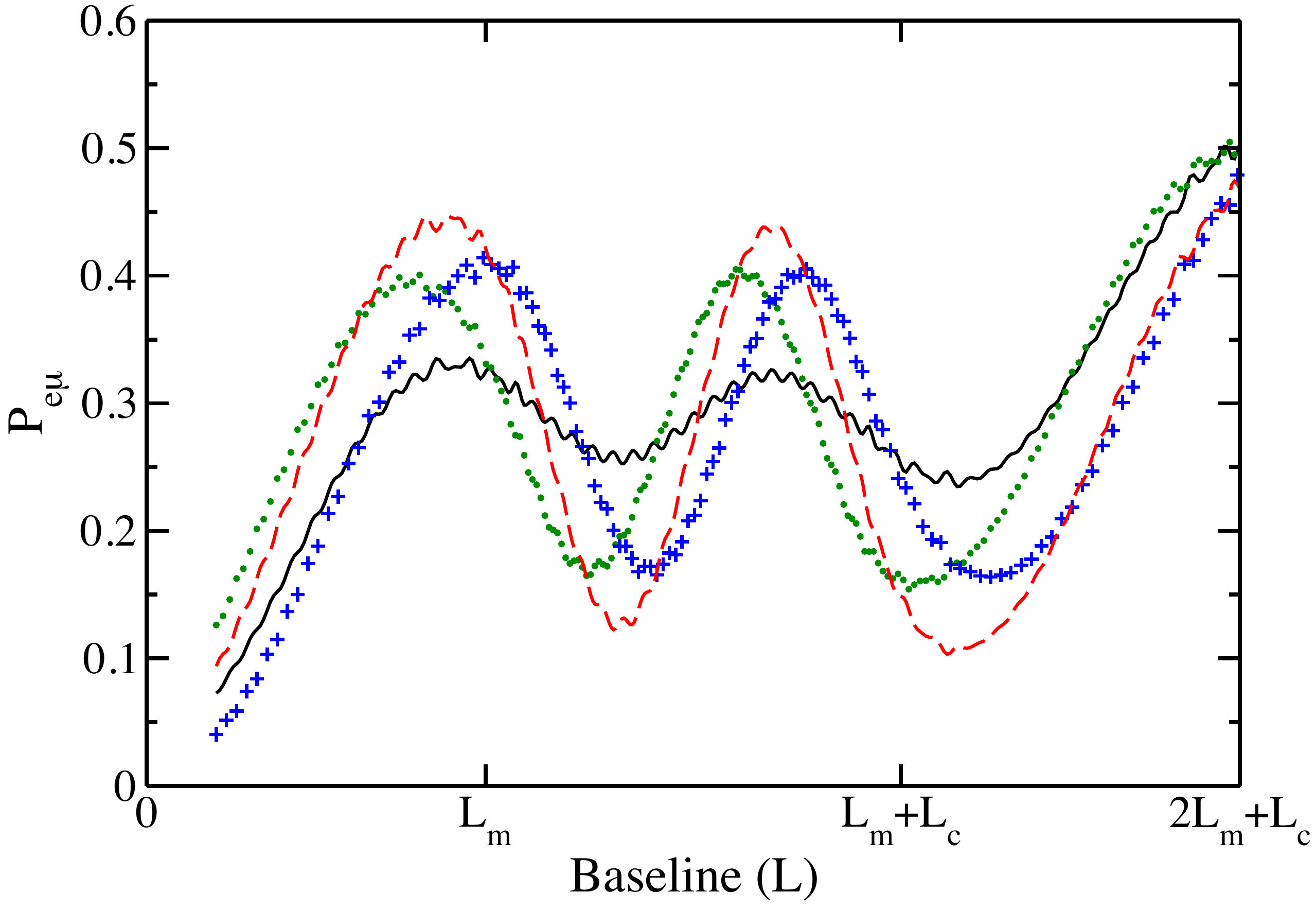}
\caption{(Color online) Oscillation probability $\nu_e \to \nu_\mu$ for a chord through the earth with $\cos \Theta = -0.905$ which corresponds to  $L_m =3576$ km and $L_c = 4376$ km.  The various curves correspond to different values of the CP phase $\delta$ as in Fig.~\ref{fig4}. 
\label{fig9} }
\end{figure}

\begin{figure}[th]
\includegraphics[width=8.6cm]{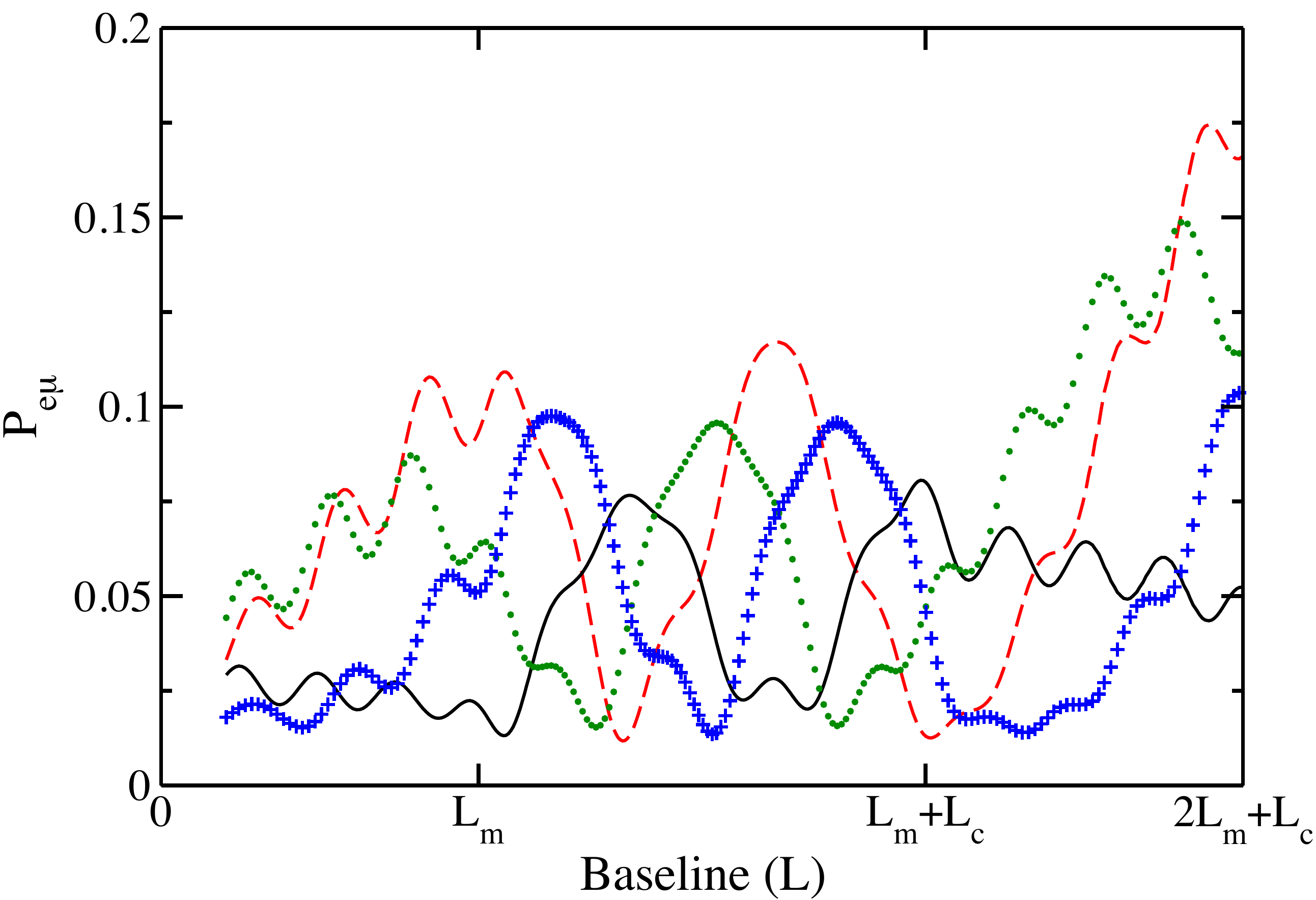}
\caption{(Color online) Oscillation probability $\nu_e \to \nu_\mu$ for a chord through the earth with $\cos \Theta = -0.919$ which corresponds to  $L_m =3438$ km and $L_c = 4834$ km.  The various curves correspond to different values of the CP phase $\delta$ as in Fig.~\ref{fig4}. \label{fig10} }
\end{figure}

To explore this point further, we plot in Figs.~\ref{fig9} and \ref{fig10} the $\nu_e \to \nu_\mu$ oscillation probability for the 200 MeV and 800 MeV neutrinos for paths along the chords which correspond to parametric resonance,  $\cos \Theta=-0.905$ and $-0.919$, respectively.  As the terminus of the neutrino's path through the earth is not  an integer number of periods, we cannot use Eq.~(\ref{pem_3nu}) to ascertain $\mathcal{P}_{e\mu}$ here. At best, the previous analysis only informs our knowledge of the state as the neutrino leaves the core at $L=L_m+L_c$; from Eq.~(\ref{state_kl}), the state, in the propagation basis, is
\begin{equation}
\tilde{\nu}(L)  \approx \left( \begin{array}{c}
\cos \Phi\\
-ie^{i \gamma} \sin\Phi\\
\theta_{13} \tilde{\mathcal{U}}_{\tau \tau}(L)
\end{array}
\right) \label{nu_earth}
\end{equation}
where the phases $\Phi$ and $\gamma$ can be read off of the plots in Fig.~\ref{fig7}.  For the remaining bit of the path through the mantle, one can simply operate on this state with a constant density evolution operator $\tilde{\mathcal{U}}^A(L_m)$, Eq.~(\ref{time_ev_prop}), to determine the state at the end of the chord.  


Though the curves in Figs.~\ref{fig9} and \ref{fig10} are rather different, they can be understood within the same framework as both systems approximately satisfy the half-wavelength condition with $\sin \varphi_m =1 = \sin \varphi_c$.  A consequence of the half-wavelength is the fact that the phase $\gamma$ takes the value $\pm \frac{\pi}{2}$.  For both the 200 MeV and 800 MeV trajectories, this phase is actually $\gamma \simeq -0.4 \pi$; however, we will assume $\gamma = -\pi/2$.  Given this, we deduce from Eq.~(\ref{xvec}) that $\sin \Phi= \sin(2\theta_c - 2\theta_m)$.  With this simplification along with the assumption for $\gamma$, we find that for the half-wavelength condition the oscillation probability at the end of the neutrino's trajectory is
\begin{eqnarray}
\mathcal{P}_{e\mu}(2L_m+L_c)& \approx& c_{23}^2\sin^2(4\theta_m-2\theta_c)\nonumber \\
&&+\theta_{13}s_{23}c_{23}c_\delta \sin(8\theta_m-4\theta_c). 
\end{eqnarray}
As 800 MeV is roughly eight times the MSW resonance energy in the mantle, the overall amplitude of these oscillations is suppressed relative to the 200 MeV case. 
So far as the half-wavelength condition is satisfied, the CP phase only enters in the second term of the righthand side of the previous equation.  As it enters only as $\cos \delta$, there will be no difference between curves with $\delta = \frac{\pi}{2}, \frac{3\pi}{2}$. The difference between the  CP conserving and maximally violating cases is given by
\begin{equation}
\mathcal{P}_{e\mu_{\delta = 0}} - \mathcal{P}_{e\mu_{\delta = \frac{\pi}{2}}} \approx \theta_{13}s_{23}c_{23} \sin(8\theta_m-4\theta_c).
\end{equation}
For a particular matter profile, the size of this difference is controlled by the factor involving the effective $\theta_{12}$ mixing angles in the mantle and core.  For the 200 MeV case, one has $\sin(8\theta_m-4\theta_c) \approx 0.31$, and for the 800 MeV neutrinos, this is $\sin(8\theta_m-4\theta_c) \approx -0.81$.   This accounts for the large separation between the curves in Fig.~\ref{fig10}, relative to the 200 MeV case.  In terms of maximizing this difference, $|\sin(8\theta_m-4\theta_c)|=1$ for $E=500$ MeV, so in principle, this energy would be best to differentiate the CP conserving or violating cases for atmospheric neutrinos traveling through the earth along a chord which satisfies the half-wavelength condition.

\begin{figure}[th]
\includegraphics[width=8.6cm]{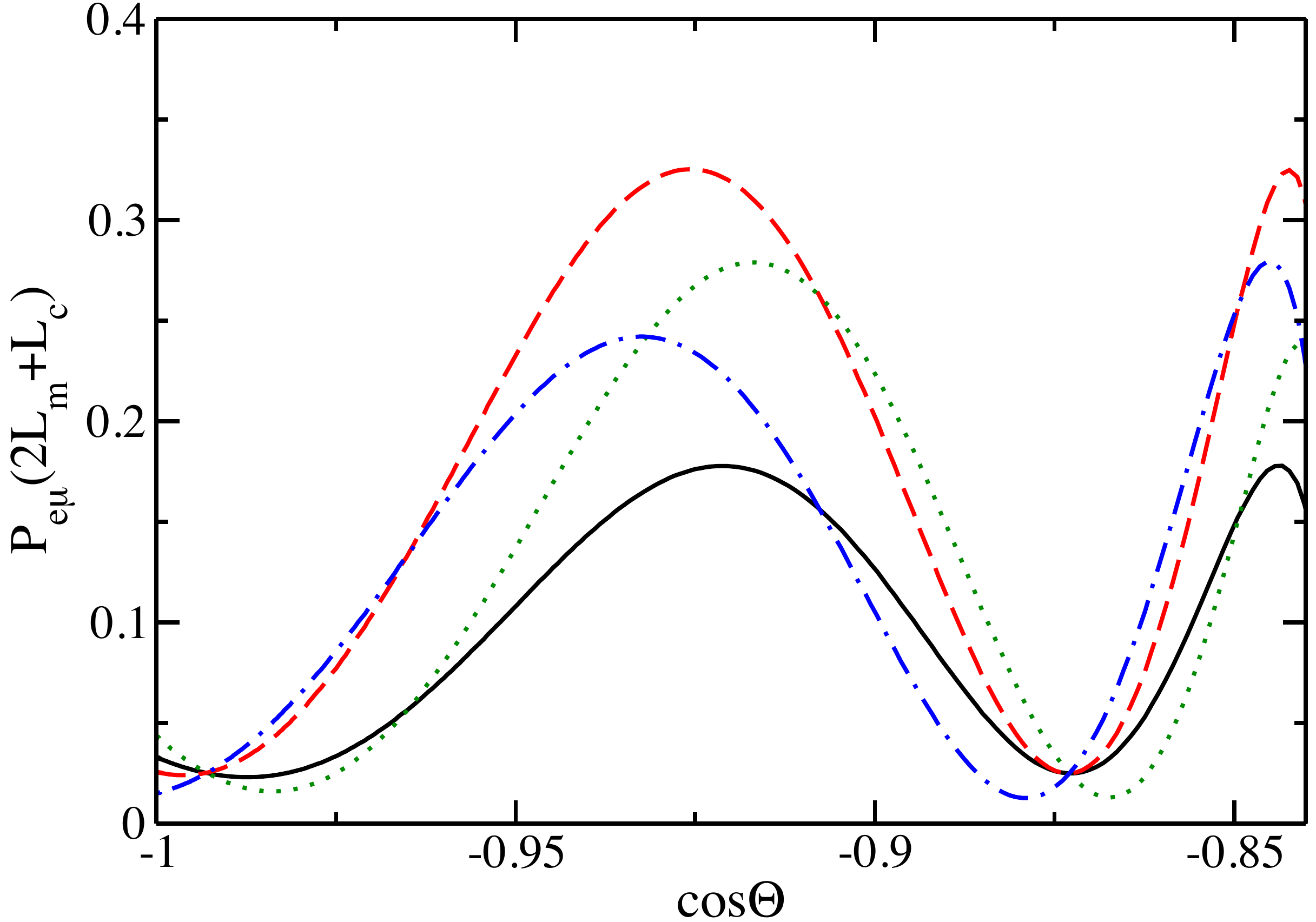}
\caption{(Color online) The oscillation probability $\mathcal{P}_{e\mu}$ for a neutrino which has traveled along a chord through the earth with zenith angle $\Theta$ averaged over a flat neutrino spectrum from 400 MeV to 600 MeV. The various curves correspond to different values of the CP phase $\delta$ as in Fig.~\ref{fig8}. \label{fig11} }
\end{figure}

In an actual experiment, the finite energy resolution of the detector must be considered; fortunately, the effects discussed herein are not washed out by a broad spectrum neutrino source.  Focusing upon the optimal energy to ascertain CP violation, 500 MeV, we consider a neutrino beam with a flat energy spectrum between 400 MeV and 600 MeV.  In Fig.~\ref{fig11}, we plot the beam's oscillation probability $\mathcal{P}_{e\mu}$ along a chord through the earth parametrized by the zenith angle $\Theta$.  The curves which have conserved CP symmetry, $\delta=0,\pi$, and maximal CP violation, $\delta = \frac{\pi}{2},\frac{3\pi}{2}$, should be experimentally discernible for this broadband source.  

The bulk of current data on atmospheric neutrinos comes from the Super-Kamiokande collaboration \cite{skatm1,skatm2,skatm3,skatm4}.  Unfortunately, this data cannot be used to search for the effects discussed in this paper because, most crucially, a water Cerenkov detector cannot distinguish between neutrino and anti-neutrino events and, secondarily, the ability to correctly ascertain the incident neutrino's zenith angle from the (detected) charged lepton is poor below 1 GeV.  A proposed detector, the magnetized iron calorimeter (ICAL), at the India Neutrino Observatory (INO) \cite{ino_cpv} can distinguish muon neutrinos from anti-neutrinos,  overcoming the primary impediment of a water Cerenkov detector.   The resolution in zenith angle for the ICAL is still poor at low energies \cite{samanta_hier}.  If we were to include this additional uncertainty in our computations, the peaks in Fig.~\ref{fig11} will become smeared out, but the curves should still be experimentally distinguishable.

\section{Conclusion}

Using an approximation appropriate for sub-GeV neutrinos traveling through the earth, we are able to study  in a three-neutrino framework parametric resonance of neutrino oscillations for a periodic density profile.  Commensurate with the initial level of approximation, we develop a parametric resonance condition similar to the exact condition for two-neutrino systems.  For a castle wall density profile, it is shown that the $\nu_e \to \nu_\mu$ oscillation probability at an integer number of periods is enhanced and bounded by $\cos^2 \theta_{23}$.  The CP phase $\delta$ enters into the oscillation probability at these points via a phase $\psi$ which is proportional to $\theta_{13}$ and involves the phase $\gamma$, a characteristic of the density profile and neutrino energy.  This phase is present in an exact  two-neutrino framework of parametric resonance but is of no measurable consequence in that context.

As expected, in the three-neutrino framework parametric resonance significantly enhances the oscillation probability.  We examine in detail instances of parametric resonance in which the phase $\gamma$ takes on three different values.  When $\gamma=\frac{\pi}{2}$ , the $\nu_e\to\nu_\mu$ oscillation probabilities achieve maximal separation at the end the first period for $\delta=0$ and $\delta=\pi$, provided $\Phi = \frac{\pi}{4}$.   When $\gamma=\frac{\pi}{4}$, this oscillation probability achieves maximal separation for the $\delta=0$ and $\delta=\frac{\pi}{2}$ cases, given the same condition for $\Phi$.  Though these trajectories are best for differentiating $\delta=0$ and $\delta=\frac{\pi}{2}$, they also suffer degeneracies for $\delta=0,\frac{3\pi}{2}$ and $\delta=\frac{\pi}{2},\pi$.
Finally, whenever $\gamma=0$ and $\Phi = \frac{\pi}{2}$, the $\mathcal{P}_{e\mu}$ oscillation probability for all values of $\delta$ is the same an the end of each period.

We also apply this formalism to sub-GeV neutrinos which travel along a chord through the earth, using the mantle-core transition to generate parametric resonance.  Significant enhancement of the oscillation probability exists even in the case in which the neutrino energy is far from the MSW resonance.  Though a path through the earth is not an integer number of periods, the formalism is useful to determine the state of the neutrino upon leaving the core. As the trajectories through the earth nearly satisfy the half-wavelength condition, the oscillation formulae simplify greatly.  Insofar as this condition is satisfied, we note that energies near 500 GeV will be best for differentiating the $\delta=0$ and $\delta=\frac{\pi}{2}$ cases in the $\nu_e \to \nu_\mu$ oscillation channel.

As for a clean experimental confirmation of these resonant oscillations, two main impediments exist.  First, the ability to differentiate between neutrino and anti-neutrino events is crucial.  Water Cerenkov detectors like Super-K do not have this ability, but ICAL at INO could overcome this.  The second issue deals with the ability to accurately assess a neutrino's incoming zenith angle $\Theta$.  For sub-GeV neutrinos, the ICAL detector has poor resolution for the zenith angle, but at 500 GeV, a detector should still be able to distinguish the cases of $\delta=0,\pi$ and $\delta=\frac{\pi}{2},\frac{3\pi}{2}$.  Regardless, a semi-analytic understanding of the interplay between parametric resonances and CP violation for sub-GeV neutrinos provides a useful backdrop for future data analysis.

\section{ACKNOWLEDGMENTS}

We thank B.~K.~Cogswell whose comments improved the clarity of this manuscript.
The work of E.~A.~H.~was supported, in part, by the Dean's Summer Research Grant from Reed College.

\bibliography{biblio}

\end{document}